\documentclass[sn-mathphys]{sn-jnl}
\newcommand{\showprivate}[1]{}

\newif\iffull
\fulltrue

\usepackage{amsmath}
\usepackage{amsthm}
\usepackage{amssymb}
\usepackage{graphicx}
\usepackage{hyperref}
\usepackage[capitalize]{cleveref}
\usepackage{caption}
\usepackage{subcaption}
\usepackage{xcolor}
\usepackage{enumerate}
\usepackage{bbm}
\usepackage{xstring}

\newtheorem{theorem}{Theorem}
\newtheorem{lemma}{Lemma}
\newtheorem{definition}{Definition}
\newtheorem{corollary}{Corollary}

\newcommand{\cW}{\mathcal{W}}
\newcommand{\cF}{\mathcal{F}}
\newcommand{\rem}{\mathrm{rem}}

\makeatletter
\newtheorem*{rep@theorem}{\rep@title}
\newcommand{\newreptheorem}[2]{%
\newenvironment{rep#1}[1]{%
 \def\rep@title{#2 \ref{##1}}%
 \begin{rep@theorem}}%
 {\end{rep@theorem}}}
\makeatother

\newreptheorem{theorem}{Theorem}
\newreptheorem{lemma}{Lemma}
\newcommand{\req}{\textsc{req}}

\newcommand{\lookupauthors}[1]{%
    \IfEqCase{#1}{%
        {loulou1973multi}{Loulou}%
        {kollerstrom1974heavy}{Köllerström}%
        {kollerstrom1979heavy}{Köllerström}%
        {goldberg2017simple}{Goldberg and Li}%
        {van2003markovian}{Van Harten and Sleptchenko}%
        {boxma2002waiting}{Boxma et al.}%
        {gupta_2009_adaptive}{Gupta and Harchol-Balter}%
        {telek_response_2018}{Telek and Van Houdt}%
        {berg_towards_2017}{Berg et al.}%
        {berg_optimal_2020}{Berg et al.}%
        {sigman1994finite}{Sigman and Yao}%
        {scheller_finite_1996}{Scheller-Wolf}%
    }[\GenericError{}{unknown key in lookupauthors}{}{add it to the big IfEqCase}\textbf{??}]}
\newcommand{\citemanual}[1]{\lookupauthors{#1}{~\cite{#1}}}

\jyear{2022}

\begin{document}
\title{WCFS: A new framework for analyzing multiserver systems}
\abstract{
    Multiserver queueing systems are found at the core of a wide variety of practical systems.
    Many important multiserver models have a previously-unexplained similarity:
    identical mean response time behavior is empirically observed in the heavy traffic limit.
    We explain this similarity for the first time.

    We do so by introducing the work-conserving finite-skip (WCFS) framework,
    which encompasses a broad class of important models.
    This class includes 
    the heterogeneous M/G/k, the limited processor sharing policy for the M/G/1,
    the threshold parallelism model, and the multiserver-job model
    under a novel scheduling algorithm.

    We prove that for all WCFS models,
    scaled mean response time $E[T](1-\rho)$ converges to the same value, $E[S^2]/(2E[S])$,
    in the heavy-traffic limit,
    which is also the heavy traffic limit for the M/G/1/FCFS.
    Moreover, we prove additively tight bounds on mean response time
    for the WCFS class, which hold for all load $\rho$.
    For each of the four models mentioned above,
    our bounds are the first known bounds on mean response time.
}

\author*[1]{\fnm{Isaac} \sur{Grosof}}\email{igrosof@cs.cmu.edu}

\author[1]{\fnm{Mor} \sur{Harchol-Balter}}\email{harchol@cs.cmu.edu}

\author[2]{\fnm{Alan} \sur{Scheller-Wolf}}\email{awolf@andrew.cmu.edu}

\affil[1]{
    \orgdiv{Computer Science Department},
    \orgname{Carnegie Mellon University},
    \orgaddress{
        \street{5000 Forbes Ave},
        \city{Pittsburgh},
        \postcode{15213},
        \state{PA},
        \country{USA}%
    }%
}
\affil[2]{
    \orgdiv{Tepper School of Business},
    \orgname{Carnegie Mellon University},%
    \orgaddress{
        \street{5000 Forbes Ave},
        \city{Pittsburgh},
        \postcode{15213},
        \state{PA},
        \country{USA}%
    }%
}


\maketitle
\section{Introduction}
Consider the following four queueing models,
which are each important, practical models,
but which seem very different.
We will refer to these models throughout the paper
as our \emph{four motivating models}:
\begin{itemize}
    \item \textbf{Heterogeneous M/G/k:} A $k$-server system where servers run
    at different speeds.
    Jobs are held at a central queue and served in First-Come-First-Served (FCFS)
    order when servers become available.
    If multiple servers are vacant,
    a server assignment policy
    such as Fastest Server First is applied.

    \item \textbf{Limited processor sharing:}
    A single-server system 
    where if at least $k$ jobs are present,
    the $k$ earliest arrivals each receive an equal fraction of service.
    If fewer than $k$ jobs are present,
    the server is split equally among all jobs.

    \item \textbf{Threshold parallelism:}
    A multiserver system where jobs can run on any number of servers up to some threshold,
    with perfect speedup.
    We consider FCFS service, where each job is allocated a number of servers
    equal to its threshold,
    as long as servers are available.
    The final job served may be allocated fewer servers than its threshold. 

    \item \textbf{Multiserver-jobs under the ServerFilling policy:}
    A multiserver system
    where the jobs are called ``multiserver jobs,''
    because each job requires a fixed number of servers,
    which it holds concurrently throughout its service.
    We examine a service policy called \textit{ServerFilling},
    which always fills all of the servers if enough jobs are available.
\end{itemize}
We define these models in more detail in \cref{sec:special-models}.

\begin{figure}[t]
\includegraphics[width=\textwidth]{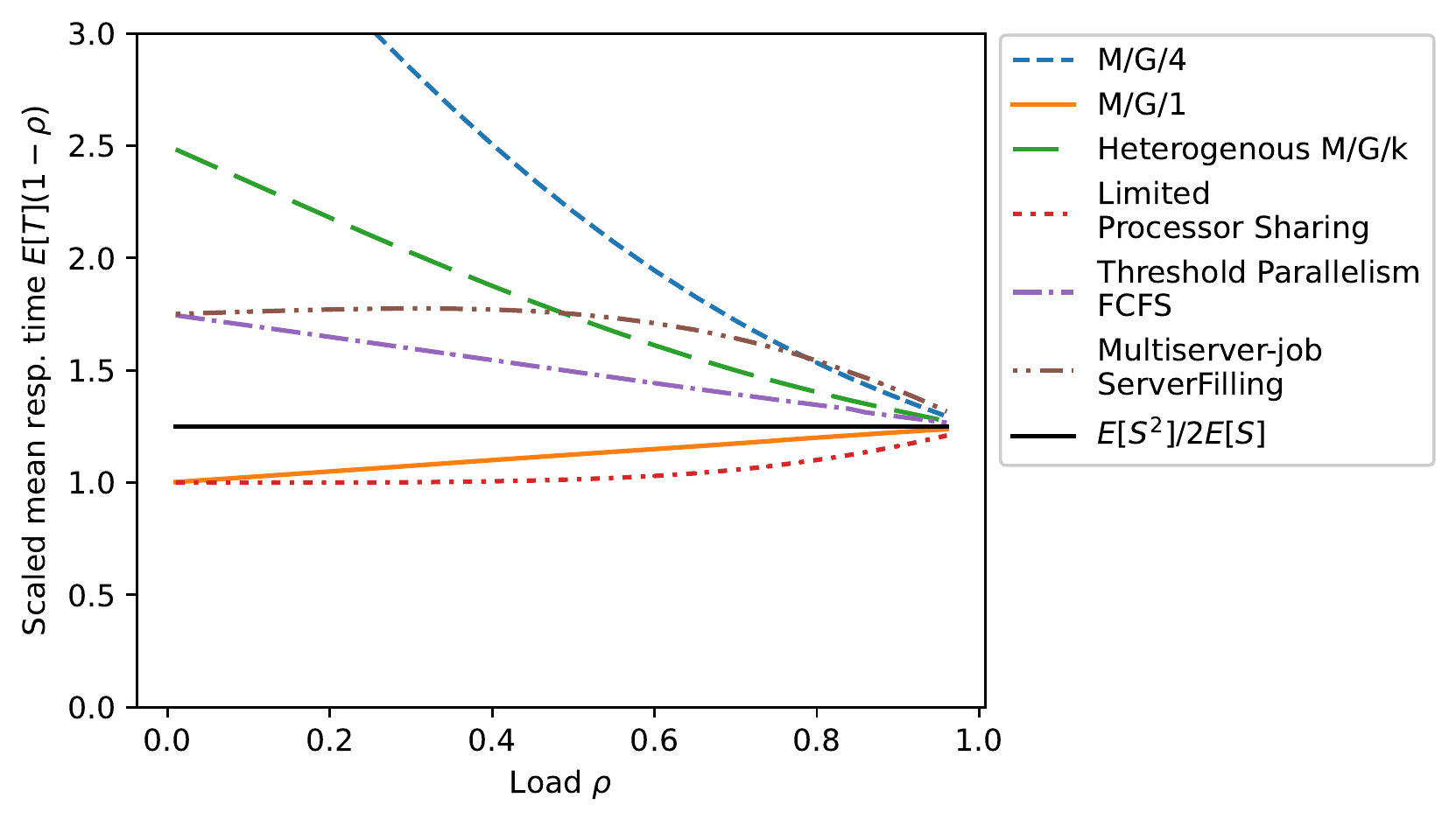}
\caption{Scaled mean response time of our four motivating models,
as well as the related M/G/k and M/G/1 models.
Our four motivating models will be further defined in \cref{sec:special-models}.
In each case, the job size distribution $S$
is distributed as $Hyperexp(\mu_1 = 2, \mu_2 = \frac{2}{3}, p_1 = \frac{1}{2})$.
The black line is $E[T](1-\rho) = \frac{E[S^2]}{2E[S]}$,
the heavy traffic behavior of M/G/1/FCFS and each of our models of interest.
$10^9$ arrivals simulated. $\rho \in [0, 0.96]$ to ensure accurate results.}
\label{fig:intro-response-wcfs}
\end{figure}

We will show that, while our four motivating models appear quite different,
their mean response times, $E[T]$, are very similar,
especially in the heavy-traffic limit.
Specifically, we will show that
their behavior in the heavy traffic limit is identical
to that of the M/G/1/FCFS model,
and in fact the mean response time
of each of these disparate models
only differs by an \emph{additive} constant from that of M/G/1/FCFS for all loads,
a much stronger result
than convergence
in heavy traffic.

The similarity of these models is illustrated by \cref{fig:intro-response-wcfs},
which shows
mean response time, $E[T]$, scaled by a factor of $1-\rho$,
to help illustrate the asymptotic behavior in the $\rho \to 1$ limit.
Observe that in each of our models of interest,
as well as in the M/G/1 and the M/G/4,
$E[T](1-\rho)$
converges to $E[S^2]/2E[S]$, the mean of the equilibrium (excess) distribution,
where $S$ denotes the job size distribution and $\rho = \lambda E[S] < 1$ is the system load.

\begin{figure}
\includegraphics[width=\textwidth]{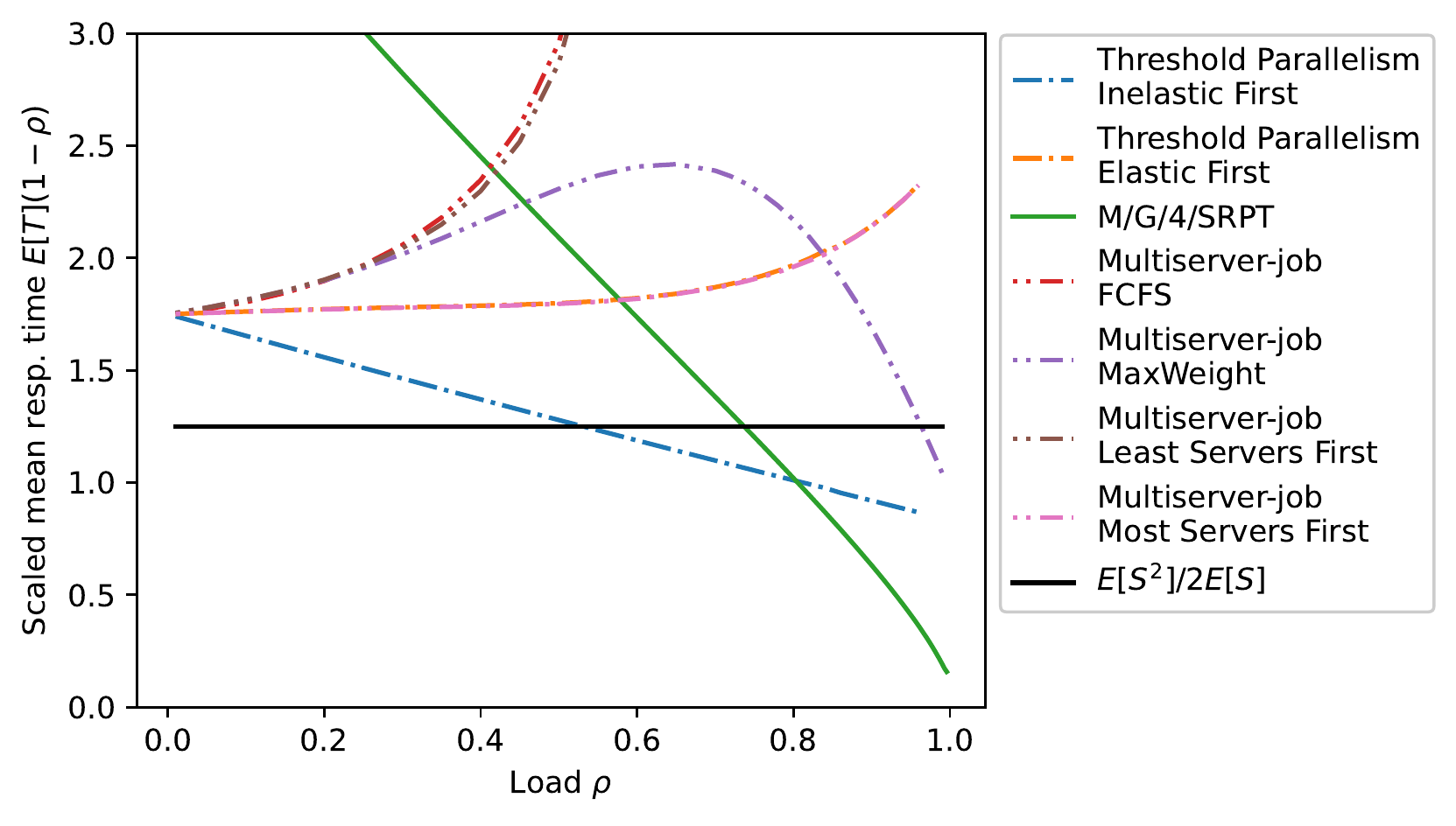}
\caption{Scaled mean response time of alternative models and policies.
All of these models and policies will be explained in \cref{sec:empirical}.
$S \sim Hyperexp(\mu_1 = 2, \mu_2 = \frac{2}{3}, p_1 = \frac{1}{2})$.
Black line is $E[T](1-\rho) = \frac{E[S^2]}{2E[S]}$.
$10^9$ arrivals simulated, $\rho \in [0, 0.96]$ to ensure accurate results,
except MaxWeight and M/G/4/SRPT: $10^{10}$ arrivals, $\rho \in [0, 0.99]$.}
\label{fig:intro-response-other}
\end{figure}

This similarity is striking --
to see just how notable it is, consider a variety of alternative models and policies
shown in \cref{fig:intro-response-other}.
For these alternative models, scaled mean response time either does not converge at all,
or converges to a different limit entirely.

This contrast poses an intriguing question:
\begin{quote}
    \textit{Why do our four motivating models converge to M/G/1/FCFS in heavy traffic?}
\end{quote}
To put it another way,
we ask what crucial property our four motivating models share,
that is not shared by the alternative models in \cref{fig:intro-response-other}.

To answer this question,
we define the ``work-conserving finite-skip'' framework (WCFS),
which applies to a broad class of models.
The WCFS class contains our
four motivating queueing models, as well many others.
We demonstrate that for any model in the WCFS class (which we call a ``WCFS model"),
if the job size distribution $S$ has bounded expected remaining size,
then its scaled mean response time converges to the same heavy traffic limit as the M/G/1/FCFS.
Specifically, we prove that
\begin{reptheorem}{thm:scaled-et}
    For any model $\pi \in$ WCFS with bounded expected remaining size\footnote{%
    This assumption is defined in \cref{sec:finite-rem-sup}.%
    },
\begin{align*}
    \lim_{\rho \to 1} E[T^\pi] (1-\rho) = \frac{E[S^2]}{2E[S]}.
\end{align*}
\end{reptheorem}

\cref{thm:scaled-et} follows from an even stronger result: We prove that the difference in mean response time
between any WCFS model and M/G/1/FCFS is bounded by an explicit additive constant,
that may depend on the specific WCFS model.
\begin{reptheorem}{thm:explicit-bounds}
    For any model $\pi \in$ WCFS with bounded expected remaining size,
    \begin{align*}
    E[T^\pi] &\le\frac{\rho}{1-\rho}\frac{E[S^2]}{2 E[S]} + c_{upper}^\pi \\
    E[T^\pi] &\ge\frac{\rho}{1-\rho}\frac{E[S^2]}{2 E[S]} + c_{lower}^\pi
    \end{align*}
for explicit constants $c_{upper}^\pi$ and $c_{lower}^\pi$ not dependent on load $\rho$.
\end{reptheorem}

\cref{thm:explicit-bounds} not only implies \cref{thm:scaled-et},
it also guarantees rapid convergence of scaled mean response time to the heavy traffic limit
specified in \cref{thm:scaled-et}.

In summary, this paper makes the following contributions:
\begin{itemize}
    \item We define the WCFS framework
    and our bounded expected remaining size assumption. (\cref{sec:model})
    \item We prove that each of the four motivating models
    is a WCFS model. (\cref{sec:special-models})
    \item We discuss prior work on WCFS models. (\cref{sec:prior})
    \item We prove that all WCFS models with bounded expected remaining size
    have the same scaled mean response time as M/G/1/FCFS,
    and mean response time within an additive constant of M/G/1/FCFS.
    (\cref{sec:results})
    \item We empirically validate our results,
    contrasting heavy traffic behavior of WCFS models and non-WCFS models.
    (\cref{sec:empirical})
\end{itemize}

\section{The WCFS Framework and WCFS Models}
\label{sec:model}

In \cref{sec:wcfs-models,sec:wcfs-examples},
we define the WCFS framework and resulting class of models.
In \cref{sec:finite-rem-sup},
we define our ``bounded expected remaining size'' assumption.
In \cref{sec:notation}, we define a few more concepts that will be used in the paper.

Job sizes are sampled $i.i.d.$ from a job size distribution.
Once sampled, job sizes are fixed:
we assume preempt-resume service if a job is preempted while in service.
Intuitively, the size of a job represents the amount of work associated with the job.
Size will be defined in more detail in \cref{sec:work-conserving}.

\subsection{WCFS Framework and WCFS Models}
\label{sec:wcfs-models}
The WCFS framework applies to the class of models with
Poisson arrivals at rate $\lambda$, which satisfy the following properties:
\begin{enumerate}
    \item Finite skip (\cref{sec:finite-skip}),
    \item Work conserving (\cref{sec:work-conserving})
    \item Non-idling (\cref{sec:positive-b-min}).
\end{enumerate}

\subsubsection{Finite skip}
\label{sec:finite-skip}
\begin{figure}
    \includegraphics[width=\textwidth]{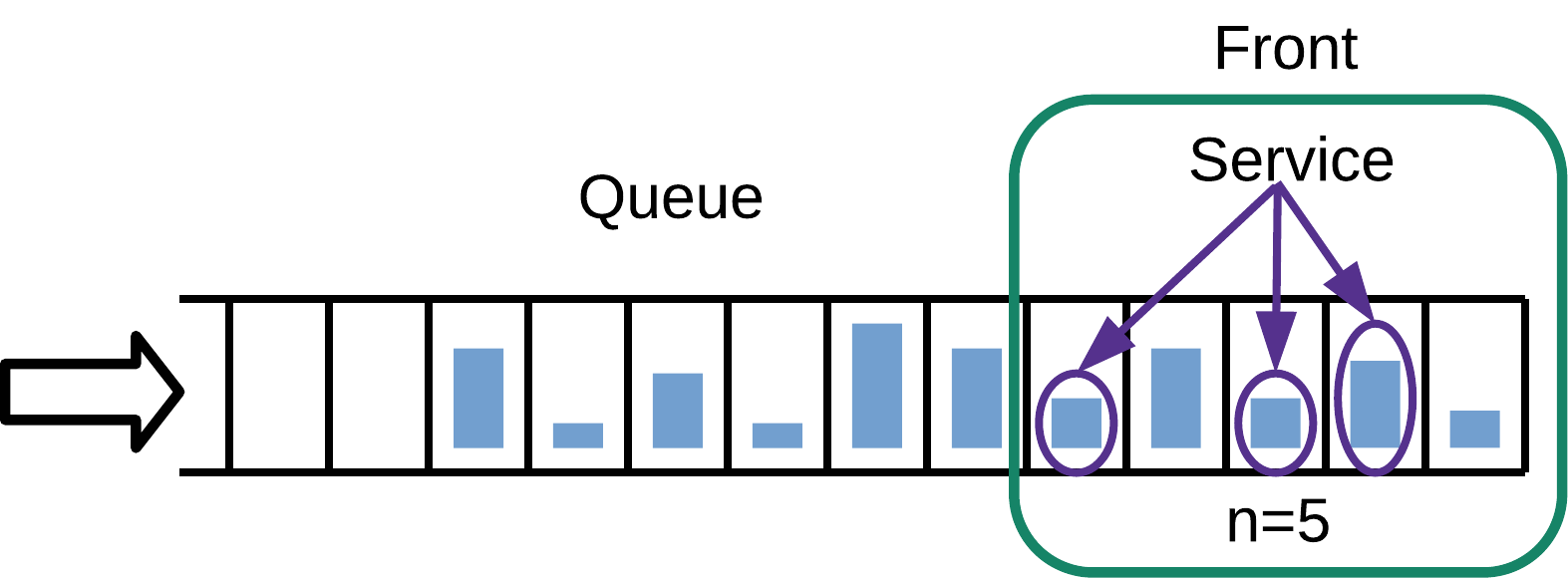}
    \caption{Diagram of a Finite-Skip Model}
    \label{fig:diagram}
\end{figure}

We first define the finite-skip property, which defines the class of finite-skip models.
Consider the jobs in the system in arrival order.
Associated with each finite-skip model,
there is a finite parameter $n$.
We partition the jobs in the system into two sets:
the (up to) $n$ jobs which arrived longest ago,
which we call the \emph{front},
and all other jobs, which we call the \emph{queue}.
The \emph{finite-skip property} specifies that,
among all of the jobs in the system,
the server(s) only serve jobs in the front.
In particular, no jobs beyond the first $n$ jobs in arrival order
receive any service.
\cref{fig:diagram} shows a generic finite-skip model.

\begin{definition}
We call the front \emph{full} if at least $n$ jobs are present in the system,
and therefore exactly $n$ jobs are at the front.
\end{definition}

The intuition behind the term ``finite skip''
comes from imagining moving through the jobs in the system in arrival order,
skipping over some jobs and serving others.
In a finite-skip model only the first $n$ jobs can be served,
so only finitely many jobs can be skipped.

\subsubsection{Work conserving}
\label{sec:work-conserving}
Now, we will specify what we mean by ``work conserving,''
which is a different concept here than in previous work.

First, we normalize the total system capacity to 1,
regardless of the number of servers in the system.
For instance, in a homogeneous $k$-server system,
we think of each server as serving jobs at rate $1/k$.

Whenever a job is in service, it receives some fraction of the system's total service capacity,
which we call the job's \emph{service rate}.
Let $B(t) \le 1$ denote the total service rate of all jobs in service at time $t$,
and let $B$ be the stationary total service rate,
assuming for now such a quantity exists.

We define a job's \emph{age} at time $t$ to be the total amount of service
the job has received up to time $t$:
a job's age
increases at a rate equal to the job's service rate
whenever the job is in service.
Each job has a property called its \emph{size}.
When the job's age reaches its size,
the job completes.

In particular, we assume that every job $j$ has a size $s_j$
and a class $c_j$
drawn i.i.d. from some general joint distribution.
Let $(S, C)$ be the random variables denoting a job's size and class pair.
A job's class is static information known to the scheduler,
while a job's size is unknown to the scheduler.
For instance, in the threshold parallelism model defined in \cref{sec:threshold-parallelism},
a job's parallelism threshold is its class.

\begin{definition}
We call the system \emph{maximally busy}
if the entire capacity of the system is in use,
namely if the total service rate of jobs in service is 1.

We define a finite-skip model to be \emph{work conserving}
if whenever the front is full,
the system is also maximally busy.
\end{definition}
In other words, a finite-skip model is work-conserving
if, whenever there are at least $n$ jobs in the system,
the total service rate is 1.

Now that we have defined a job's size,
we can also define the load of the system:
$\rho = \lambda E[S]$.
Load $\rho$ is the time-average service rate,
or equivalently the time-average fraction of capacity in use.
Specifically, $\rho = E[B]$.
We assume $\rho < 1$ to ensure stability.

\subsubsection{Non-idling}
\label{sec:positive-b-min}
We also assume that the total service rate $B(t)$ is bounded away from zero
whenever a job is present.
Specifically, whenever a job is present,
we assume that $B(t) \ge b_{\inf}$,
for some constant $b_{\inf} > 0$.

This assumption is key to bounding mean response time under low load.
For an example, see the batch-processing system in \cref{sec:wcfs-examples}.

\subsection{Examples and non-examples}
\label{sec:wcfs-examples}
To clarify which models fit within the WCFS framework,
we give several examples, both positive and negative.
\begin{itemize}
    \item \textbf{M/G/k/FCFS:} This is a WCFS model with $n=k$.
    \item \textbf{M/G/$\infty$:} This model is not finite skip.
    All jobs are in service, regardless of the number of jobs in the system:
    there is no finite bound on the number of jobs in service.

    \item \textbf{M/G/k/SRPT:} In this model,
    the $k$ jobs with smallest remaining size are served at rate $1/k$.
    This model is not finite skip because the jobs with smallest remaining size
    can be arbitrarily far back in the arrival ordering.

    \item \textbf{Multiserver-job model:}
    Consider a multiserver system with $k=2$ servers,
    and where each job requires either 1 or 2 servers.
    Let the front size $n = 2$.

    If jobs are served in FCFS order, with head-of-the-line (HOLB) blocking,
    this policy is finite-skip, but not work-conserving.
    If the front consists of a job requiring 1 server followed by a job requiring 2 servers,
    under HOLB the system will only utilize one server.
    In this case, the front is full, because $n=2$ jobs are present in the system,
    but the system is not maximally busy.

    In contrast, consider a service policy
    which serves a 2 server job if either of the jobs in the front are 2 server jobs,
    or else serves each of the 1 server jobs at the front.
    This policy is a special case of the ServerFilling policy,
    depicted in \cref{fig:intro-response-wcfs} and defined in general
    in \cref{sec:server-filling}.
    This policy is finite-skip and work-conserving.
    \item \textbf{Batch-processing M/G/k:}
    If there are at least $k$ jobs present,
    the oldest $k$ jobs in the system are each served at rate $\frac{1}{k}$.
    Otherwise, no service occurs.
    This model is finite-skip and work-conserving,
    but is not non-idling.
    To see why the non-idling property is necessary for our main results,
    specifically \cref{thm:explicit-bounds},
    one can show that 
    in the $\lambda \to 0$ limit,
    response times will grow arbitrarily large in the batch-processing M/G/k.
    To rule out systems where $E[T]$ diverges in the $\lambda \to 0$ limit,
    we assume the non-idling property.
    \item \textbf{Red and Blue M/G/k:}
    Imagine an M/G/k with red and blue jobs. Only one color of jobs is allowed to be in service at a time.
    To determine which jobs to serve, the scheduler counts off jobs in arrival order until it finds $k$ red jobs or $k$ blue jobs and serves all $k$ of the appropriate color (if fewer than $k$ jobs are found for both colors, the system serves the more populous color).  This scheduling policy is WCFS with $n = 2k - 1$.  
\end{itemize}

\subsection{Bounded expected remaining size: Finite $\rem_{\sup}$ }
\label{sec:finite-rem-sup}
At a given point in time,
let the \emph{state} of a job $j$ consist of
its class $c_j$ and its age $a_j$.
Within our WCFS framework,
we allow service to be based on the states of the jobs in the front,
but not on the number or states of jobs in the queue.

A key assumption we make is that
jobs have bounded expected remaining size from an arbitrary state.
Let $S_c$ be the job size distribution for jobs of class $c \in C$.
We define $\rem_{\sup}(S, C)$ to be the supremum over the expected remaining sizes
of jobs, taken over all states:
\begin{align*}
    \rem_{\sup}(S, C) := \sup_{c \in C, a \in \mathbb{R}^+} E[S_c - a \mid S_c > a].
\end{align*}
When size $S$ is independent of class $C$,
or when a model has no class information,
we simply write $\rem_{\sup}(S)$.

In this paper,
we focus on job size distributions
for which $\rem_{\sup}(S, C)$ is finite.
To better understand the finite $\rem_{\sup}(S, C)$ assumption,
let's walk through a couple of examples.
In all of these examples,
let's suppose that the class information is independent of the job size distribution $S$,
so we can simply write $\rem_{\sup}(S)$.

Consider a job size distribution $S$ that is hyperexponential:
\begin{align*}
    S = 
    \begin{cases}
        Exp(\mu_1) &\text{w.p. } p_1 \\
        Exp(\mu_2) &\text{w.p. } p_2 \\
        Exp(\mu_3) &\text{w.p. } p_3
    \end{cases}
\end{align*}
For all ages $a$,
the expected remaining size is bounded:
\begin{align*}
    E[S - a \mid S > a] \le \frac{1}{\min(\mu_1, \mu_2, \mu_3)} = \rem_{\sup}(S).
\end{align*}

More generally, an arbitrary phase type job size distribution $S'$
must have finite $\rem_{\sup}$.

On the other hand, Pareto job size distributions do not have finite $\rem_{\sup}$.
Let $S'' \sim Pareto(\alpha = 3, x_{\min} = 1)$, which has finite first and second moments.
\begin{align*}
    E[S'' - a \mid S'' > a] &= \frac{a}{2}, \qquad \forall a \ge 1 \\
    \lim_{a \to \infty} E[S'' - a \mid S'' > a] &= \infty \\
    \rem_{\sup} = \sup_a E[S'' - a \mid S'' > a] &= \infty
\end{align*}

In general, finite $\rem_{\sup}$ roughly corresponds to service time having
an exponential or sub-exponential tail,
though there are some subtleties.
For instance, a Weibull distribution with $P(S \ge a) = a^{-k}$ for some $k < 1$
has infinite $\rem_{\sup}$, while for $k \ge 1$, $\rem_{\sup}$ is finite.

As a final example, suppose the WCFS scheduling policy is a known-size policy,
such as a policy which serves the job with least remaining size
among the $n$ jobs in the front, at rate 1.
Because we require that service is based only on the age and class of a job,
we model this situation by saying that a job's class is its original size.
In this case, $S = C$, and the distribution $S_x$ is simply the constant $x$.
As a result, $\rem_{\sup}(S, C) = \sup(S)$.
Therefore, in a known-size setting, $\rem_{\sup}$ is finite 
only if $S$ is bounded.

\subsection{Work, Number, Response Time}
\label{sec:notation}

Let the \emph{work} in the system be
defined as the sum of the remaining sizes of all jobs in the system.
Let $W(t)$ be the total work in the system at time $t$.
Let $W_Q(t)$ and $W_F(t)$ be the work in the queue and the work at the front, respectively,
at time $t$.
(We will generally use the subscripts $_Q$ and $_F$ to denote the queue and the front.)
Let $W, W_Q,$ and $W_F$ denote the corresponding time-stationary random variables.

Recall from \cref{sec:work-conserving}
that $B(t)$ is the total service rate at time $t$.
Note that $\frac{d}{dt} W(t) = -B(t)$,
except at arrival instants.

Let $N(t)$ be the number of jobs in the system at time $t$.
Note that $N_F(t) = n$ whenever $N(t) \ge n$, because the front is full,
and $N_F(t) = N(t)$ otherwise.

Let $T$ be a random variable denoting a job's time-stationary response time:
the time from when a job arrives to when it completes.

\section{Important WCFS Models}
\label{sec:special-models}
\iffull
Here we define in more detail the four motivating models mentioned in the introduction
and depicted in \cref{fig:intro-response-wcfs},
and show that each is a WCFS model.
\fi
\subsection{Heterogeneous M/G/k}
\label{sec:heterogeneous-mgk}
The heterogeneous M/G/k/FCFS models multiserver systems where
servers have different speeds.
This scenario commonly arises in datacenters,
which are often composed of servers with a wide variety of different types of hardware
\cite{nathuji_exploiting_2007,mars_heteogeneity_2011}.
In the mobile device setting, the big.LITTLE architecture employs heterogeneous processors
to improve battery life \cite{cho2012benefits}.

Let each server $i$ have speed $v_i > 0$,
scaled so that $\sum_i v_i = 1$.
While a job is being served by server $i$,
the job's age increases at a rate of $v_i$.

If there are multiple servers idle when a job arrives,
a server is chosen according to an arbitrary server assignment policy.
Jobs may also be migrated between servers
when a job completes.
We only assume that jobs are served in FCFS order,
and that no job is left waiting while a server is idle.
Under these assumptions, all assignment policies fit within the WCFS framework.

As an example, in \cref{fig:intro-response-wcfs}
we show the scaled mean response time of a heterogeneous M/G/4
with server speeds $0.4, 0.3, 0.2, 0.1$,
and the Preemptive Fastest Server First assignment policy.

\iffull
\subsubsection{Heterogeneous M/G/k is a WCFS model}
To show that the heterogeneous M/G/k is a WCFS model,
we must verify the three properties from \cref{sec:finite-skip,sec:work-conserving,sec:positive-b-min}.
\begin{description}
    \item[Finite skip:] Jobs enter service in FCFS order.
    As a result, the jobs in service are exactly the (up to) $k$ oldest jobs in the system.
    The model is finite skip with parameter $n=k$.
    \item[Work conserving:] The system has total capacity $\sum_i v_i = 1$.
    Whenever at least $k$ jobs are present in the system,
    all servers are occupied, and the total service rate is 1.
    In other words, whenever the front is full, the system is maximally busy.
    \item[Positive service rate when nonempty:] If a job is present,
    the job will be in service on some server.
    The system will therefore have minimum service rate $b_{\inf} \ge v_{\min}$,
    where $v_{\min} = \min_i v_i$.
\end{description}
\fi

\subsection{Limited Processor Sharing}
\label{sec:limit-processor-sharing}
    The Processor Sharing policy for the M/G/1
    is of great theoretical interest,
    and has been extensively studied \cite{yashkov2007processor}.
    However, in real systems,
    running too many jobs at once causes a significant overhead.
    A natural remedy is to utilize a 
    policy is known as Limited Processor Sharing (LPS)
    \cite{nuyens2008monotonicity,telek_response_2018,zhang2008steady,gupta_2009_adaptive}.

    The LPS policy is parameterized by some Multi-Programming Level $k$.
    If at most $k$ jobs are present in the system,
    then the policy is equivalent to Processor Sharing, serving all jobs
    at an equal rate, with total service rate 1.
    When more than $k$ jobs are present,
    the $k$ oldest jobs in FCFS order
    are each served at rate $1/k$.
    LPS is a WCFS model with $n=k$.

    As an example,
    in \cref{fig:intro-response-wcfs}
    we show the scaled mean response time of
    a LPS system with MPL 4.

\iffull
\subsubsection{Heterogeneous M/G/k is a WCFS model}
To show that the heterogeneous M/G/k is a WCFS model,
we must verify the three properties from \cref{sec:finite-skip,sec:work-conserving,sec:positive-b-min}.
\begin{description}
    \item[Finite skip:] Jobs enter service in FCFS order.
    As a result, the jobs in service are exactly the (up to) $k$ oldest jobs in the system.
    The model is finite skip with parameter $n=k$.
    \item[Work conserving:] The system has total capacity $\sum_i v_i = 1$.
    Whenever at least $k$ jobs are present in the system,
    all servers are occupied, and the total service rate is 1.
    In other words, whenever the front is full, the system is maximally busy.
    \item[Positive service rate when nonempty:] If a job is present,
    the job will be in service on some server.
    The system will therefore have minimum service rate $b_{\inf} \ge v_{\min}$,
    where $v_{\min} = \min_i v_i$.
\end{description}
\fi

\subsection{Threshold Parallelism}
\label{sec:threshold-parallelism}
In modern datacenters, it is increasingly common for jobs to be parallelizable
across a variety of different numbers of servers,
where the level of parallelism is chosen by the scheduler \cite{delimitrou_quasar_2014,peng_optimus_2018}.
Under Threshold Parallelism, a job $j$ has two characteristics:
its size $s_j$ and its parallelism threshold $\ell_j$,
where $\ell_j$ is some number of servers.
Job $j$ may be parallelized across up to $\ell_j$ servers, with linear speedup.
The pair $(s_j, \ell_j)$ is sampled i.i.d. from some joint distribution $(S, L)$.
Note that $\ell_j$ is the class of the job $j$.

Let $k$ be the total number of servers.
Note that $\ell_j \in [1, k]$.
If a job $j$ is served on $q \le \ell_j$ servers,
then it receives service rate $\frac{q}{k}$
and will complete after $\frac{k s_j}{q}$ time in service.
The number of servers a job is allocated can change over time,
correspondingly changing its service rate.

We focus on the FCFS service policy.
Under this policy, jobs are placed into service in arrival order
until their total parallelism thresholds sum to at least $k$,
or all jobs are in service.
Each job $j$ other than the final job in service is served by $\ell_j$ servers.
The final job in service is served by the remaining servers.
Under FCFS service, Threshold Parallelism fits the WCFS framework
with $n=k$.

As an example, in \cref{fig:intro-response-wcfs}
we show the scaled mean response time of
a Threshold Parallelism model
where the joint distribution $(S, L)$
is $(Exp(2), 1)$ with probability $\frac{1}{2}$,
and $(Exp(\frac{2}{3}), 4)$ with probability $\frac{1}{2}$,
and with FCFS service.

As a comparison, in \cref{fig:intro-response-other},
we show Threshold Parallelism models
with the same joint distribution $(S, L)$,
but with different service policies:
``Elastic First,'' prioritizing jobs with $L=1$,
and ``Inelastic First,'' prioritizing jobs with $L=4$.
These policies do not fit within the WCFS framework,
because a job may skip over an arbitrary number of jobs.

\iffull
\subsubsection{Threshold Parallelism with FCFS service is a WCFS model}
\begin{description}
    \item[Finite skip:]
        The jobs in service are the initial set of jobs in arrival order whose
        parallelism thresholds sum to at least $k$.
        This initial set can contain at most $k$ jobs,
        because every job has parallelism threshold at least 1.
        As a result, the model is finite skip with parameter $n=k$.
    \item[Work conserving:]
        Whenever jobs are present in the system
        whose parallelism thresholds sum to at least $k$,
        all servers are occupied,
        and the system is maximally busy.
        Whenever $k$ jobs are present, the system must be maximally busy.
    \item[Positive service rate when nonempty:]
        If a job is present in the system, at least one server must be occupied,
        and so the service rate is at least $1/k$. Hence $b_{\inf} \ge 1/k$.
\end{description}
\fi

\subsection{Multiserver-jobs under the ServerFilling policy}
\label{sec:multiserver-jobs}
First, we will describe the multiserver-job setting.
Then we will specify the ServerFilling policy.

\subsubsection{Multiserver-Job Setting}
\begin{figure}
    \centering
    \includegraphics[width=0.6\textwidth]{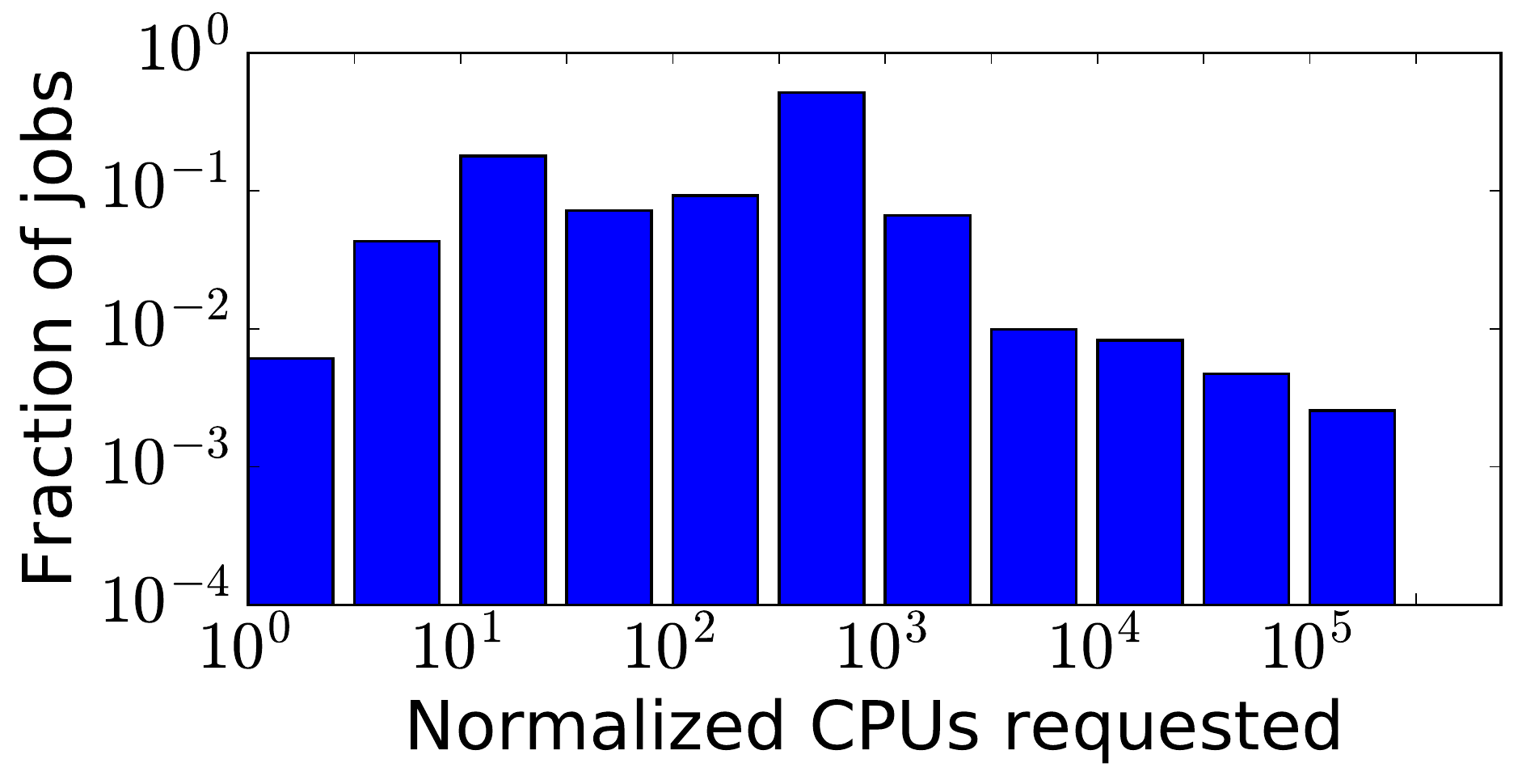}
    \caption{The distribution of number of CPUs requested
    in Google's recently published Borg trace \cite{tirmazi_borg}.
    Number of CPUs is normalized to the size of the smallest request observed,
    not an absolute value.
    }
    \label{fig:cpu_requests}
\end{figure}

When we look at jobs in cloud computing systems \cite{maguluri2012stochastic}
and in supercomputing systems \cite{feitelson_parallel_2004,srinivasan_characterization_2002,carastan_one_2019},
jobs commonly require an exact number of servers for the entire time the job is in service.
To illustrate, in \cref{fig:cpu_requests}
we show the distribution of the number of CPUs requested
by the jobs in Google's recently published trace of its ``Borg'' computation cluster
\cite{tirmazi_borg,grosof2020stability}.
The distribution is highly variable,
with jobs requesting anywhere from 1 to 100,000 normalized CPUs%
\footnote{The data was published in a scaled form \cite{tirmazi_borg}.
We rescale the data so the
smallest job in the trace uses one normalized CPU.}.

The Multiserver-Job (MSJ) model is a natural model for these computing systems.
In an MSJ model, a job $j$ has two requirements:
A number of servers $v_j$ and an amount of time $x_j$,
which are sampled i.i.d. from some joint distribution $(V, X)$.
If job $j$ requires $v_j$ servers, then it can only be served when exactly $v_j$ servers
are allocated to it. The job will complete after $x_j$ time in service.

Let a job $j$'s size be defined as
\begin{align*}
    s_j = \frac{v_j x_j}{k} \qquad S = \frac{V X}{k}.
\end{align*}

There are a wide variety of
possible service policies for placing jobs at open servers,
including FCFS, MaxWeight, Most Servers First and many others.
(We formally define these policies in \cref{sec:empirical}.)
As examples, in \cref{fig:intro-response-other},
we show the scaled mean response time of Multiserver-Job models
under a variety of service policies,
where the joint distribution $(V, X)$
is $(1, Exp(\frac{1}{2}))$ with probability $\frac{1}{2}$,
and $(4, Exp(\frac{2}{3}))$ with probability $\frac{1}{2}$.

Unfortunately, no existing policies fit within the WCFS framework --
all existing policies, including those shown in \cref{fig:intro-response-other},
are either non-finite-skip, such as Most Servers First,
or non-work-conserving, such as FCFS.
Correspondingly, in \cref{fig:intro-response-other},
we see that no existing policy has its scaled mean response time converge
to the same limit as $M/G/1/FCFS$.

We therefore define a novel service policy called ServerFilling
which yields a WCFS model.
The scaled mean response time of this service policy is depicted in \cref{fig:intro-response-wcfs},
with the same joint distribution $(V, X)$
as the policies shown in \cref{fig:intro-response-other}.
\subsubsection{ServerFilling}
\label{sec:server-filling}
For simplicity, we initially define the Server Filling policy for the common
situation in computer systems where all jobs require a number of servers which is a power of 2
($V$ is always a power of 2),
and where $k$ is also a power of 2.
We discuss generalizations in \cref{sec:generalized-server-filling}.

First, ServerFilling designates a candidate set $M$,
consisting of the minimal prefix (i.e. initial subset)
of the jobs in the system in arrival order
which collectively require at least $k$ servers.
If all jobs in the system collectively require fewer than $k$ servers,
then all are served.
Note that $\vert M\vert  \le k$
because all jobs require at least 1 server.

For instance, if $k=8$ and the jobs in the system require $[1, 2, 1, 1, 4, 2, 2, 1]$
servers, in arrival order (reading from left to right),
then $M$ would consist of the first 5 jobs:
$[1, 2, 1, 1, 4]$, which collectively require $9$ servers.

Next, the jobs in $M$ are ordered by their server requirements $v_j$,
from largest to smallest, tiebroken by arrival order.
Jobs are placed into service in that order until no more servers are available.
In our example, jobs requiring $4, 2, 1,$ and $1$ server(s) would be placed into service.

To show that ServerFilling fits within WCFS with $n=k$,
we must show that ServerFilling always utilizes all $k$ servers
if at least $k$ jobs are in the system.

\begin{lemma}
    \label{lem:server-filling}
    Let $M$ be a set of jobs such that $\sum_{j \in M} v_j \ge k$,
    where each $v_j = 2^i$ for some $i$ and $k = 2^{i'}$ for some $i'$.
    Label the jobs $m_1, m_2, \ldots$ in decreasing order of server requirement:
    $v_{m_1} \ge v_{m_2} \ge \ldots$.
    Then there exists some index $\ell \le \vert M\vert $
    such that
    \begin{align*}
        \sum_{j = 1}^\ell v_{m_j} = k.
    \end{align*}
\end{lemma}
\begin{proof}
    Let $\req(z)$ count the number of servers required by the first $z$
    jobs in this ordering:
    \begin{align*}
        \req(z) = \sum_{j=1}^z v_{m_j}.
    \end{align*}
    We want to show that $\req(\ell) = k$ for some $\ell$.
    To do so,
    it suffices to prove that:
    \begin{align}
        \label{eq:sub-goal-sf}
        \text{There exists no index } \ell' \text{ such that both } \req(\ell') < k \text{ and } 
        \req(\ell'+1) > k.
    \end{align}
    Equation~\eqref{eq:sub-goal-sf} states that $\req(z)$
    cannot cross from below $k$ to above $k$ without exactly equalling $k$.
    Because $\req(0) = 0$ and $\req(\vert M\vert ) \ge k$,
    $\req(\ell)$ must exactly equal $k$ for some $\ell$.

    To prove \eqref{eq:sub-goal-sf},
    let us examine the quantity $k-\req(z)$, the number of remaining servers
    after $z$ jobs have been placed in service.
    Because all $v_j$s are powers of 2,
    $k-\req(z)$ carries an important property:
    \begin{align}
    \label{eq:inductive-sf}
    k-\req(z)\text{ is divisible by }v_{m_{z+1}}
    \text{ for all }z.
    \end{align}
    We write $a \vert b$ to indicate that $a$ divides $b$.

    We will prove \eqref{eq:inductive-sf} inductively.
    For $z=0$,
    $k - \req(0) = k$.
    Because $k$ is a power of 2, and $v_{m_1}$ is a power of 2 no greater than $k$,
    the base case holds.
    Next, assume that \eqref{eq:inductive-sf}
    holds for some index $z$,
    meaning that $v_{m_{z+1}} \vert (k - \req(z))$.
    Note that $\req(z+1) = \req(z) + v_{m_{z+1}}$.
    As a result, $v_{m_{z+1}} \vert (k - \req(z+1))$.
    Now, note that $v_{m_{z+2}} \vert v_{m_{z+1}}$,
    because both are powers of 2, and $v_{m_{z+2}} \le v_{m_{z+1}}$.
    As a result, $v_{m_{z+2}} \vert (k - \req(z+1))$,
    completing the proof of \eqref{eq:inductive-sf}.

    Now, we are ready to prove $\eqref{eq:sub-goal-sf}$.
    Assume for contradiction that there does exist such an $\ell'$.
    Then $k-\req(\ell') > 0$, and $k-\req(\ell' + 1) < 0$.
    Because $\req(\ell' + 1) = \req(\ell') + v_{m_{\ell'+1}}$,
    we therefore know that
    $v_{m_{\ell'+1}} > k-\req(\ell')$.
    But from \eqref{eq:inductive-sf},
    we know that $v_{m_{\ell'+1}}$ divides $k-\req(\ell')$, which is a contradiction.
\end{proof}

\iffull
\subsubsection{ServerFilling for the Multiserver-Job system is a WCFS policy}
\begin{description}
    \item[Finite skip:]
        The jobs in service are a subset of the candidate set $M$,
        the initial set of jobs in arrival order whose
        server requirements $v_j$ sum to at least $k$.
        This initial set must contain at most $k$ jobs,
        because every job requires at least 1 server.
        As a result, the model is finite skip with parameter $n=k$.
    \item[Work conserving:]
        By \cref{lem:server-filling},
        whenever jobs are present in the system
        whose server requirements $v_j$ sum to at least $k$,
        all servers are occupied,
        and the system is maximally busy.
        Thus, whenever $k$ jobs are present, the system must be maximally busy.
    \item[Positive service rate when nonempty:]
        If a job is present in the system, at least one server must be occupied,
        and so the service rate is at least $1/k$. Hence $b_{\inf} \ge 1/k$.
\end{description}
\fi
\subsubsection{Generalizations of ServerFilling}
\label{sec:generalized-server-filling}
The ServerFilling policy can be generalized,
as long as 
all server requirements divide $k$.
We describe the corresponding scheduling policy,
which we call DivisorFilling,
\iffull
in \cref{app:divisor-filling}.
\fi

DivisorFilling is the most general possible WCFS policy for the MSJ setting.
If some server requirement does not divide $k$,
then no policy fits within the WCFS framework,
because the system is not work conserving if all jobs present
require that non-divisible number of servers and more than $n$ jobs are present.

\section{Prior Work}
\label{sec:prior}
\subsection{M/G/k}

\subsubsection{Fixed $k$}
In this regime, the best known bounds on response time
either require much stronger assumptions
on the job size distribution $S$ than we assume \cite{loulou1973multi},
or prove much weaker bounds on mean response time \cite{kollerstrom1974heavy,kollerstrom1979heavy}.

A paper by \citemanual{loulou1973multi} bounds mean work in system
in the M/G/k to within an additive gap,
under the strong assumption that the job size distribution $S$ is bounded.
While the paper mostly focuses on the overload regime ($\rho > 1$),
their
equations (9) and (10) apply in our setting ($\rho < 1$) as well.
They couple the multiserver system
with a single-server system on the same arrival sequence.
They show that
\begin{align*}
    0 \le W^{M/G/k}(t) - W^{M/G/1}(t) \le k \max_{1 \le i \le A(t)} S_i,
\end{align*}
where $A(t)$ is the number of jobs that have arrived by time $t$.
In the case of a bounded job size distribution $S$,
one can therefore show that
\begin{align}
    \label{eq:loulou_work}
    0 \le W^{M/G/k}(t) - W^{M/G/1}(t) \le k \sup(S).
\end{align}
One could then use this workload bound to prove a bound on mean response time in the M/G/k.

These bounds are comparable to those in our \cref{lem:work-bounds} when $S$ is bounded,
but our bounds require a much weaker assumption on the job size distribution $S$.

\citemanual{kollerstrom1974heavy} proves convergence of queueing time to an exponential distribution
in the
GI/GI/k.
Specialized to the M/G/k,
the result states that in the $\rho \to 1$ limit,
$T_Q^{M/G/k}$ converges to an exponential distribution
with mean
\begin{align*}
    \frac{\rho}{1-\rho} \frac{E[S^2]}{2 E[S]} - \frac{1}{\lambda} = E[T_Q^{M/G/1}] - \frac{1}{\lambda}.
\end{align*}

\citemanual{kollerstrom1979heavy} improves upon \cite{kollerstrom1974heavy}
by characterizing the rate of convergence
and thereby derives explicit moment bounds.
However, unlike prior single-server results \cite{kingman1962some},
these bounds are quite weak.
Specialized to the M/G/k,
\citemanual{kollerstrom1979heavy}'s bounds state that
\begin{align}
    \label{eq:k_lb}
    E[T_Q^{M/G/k}] - E[T_Q^{M/G/1}] &\ge \frac{c_{lower}}{(1 - \rho)^{1/2}} \\
    \label{eq:k_ub}
    E[T_Q^{M/G/k}] - E[T_Q^{M/G/1}] &\le \frac{c_{higher}}{1-\rho}
\end{align}
for constants $c_{lower}, c_{higher}$ not dependent on $\rho$.

The $\Theta(\frac{1}{1-\rho})$ scaling in \eqref{eq:k_ub}
is especially poor: this bound is too weak to give
any explicit bound on the convergence rate
of $E[T_Q^{M/G/k}](1-\rho)$ to the previously established limit of $\frac{E[S^2]}{2 E[S]}$.

Our bounds are tighter in that they are constants not depending on $\rho$,
but we assume $S$ has finite $\rem_{\sup}$,
while \citemanual{kollerstrom1979heavy} merely assumes that $S$ has finite second moment.

\subsubsection{Scaling $k$}

Recent work has focused on regimes where both $\rho$ and $k$ scale asymptotically,
such as the Halfin-Whitt regime.
These results are not directly comparable to ours;
they indicate
that the limiting behavior in the Halfin-Whitt regime
depends in a complex way on the job size distribution $S$ \cite{gamarnik_steady_2008,aghajani2020limit,dai2014validity}.

Turning to the more general case of scaling $k$,
in work currently under submission,
\citemanual{goldberg2017simple}
prove the first bounds on $E[T_Q]$ that scale as $\frac{c}{1-\rho}$
for an explicit constant $c$ and arbitrary joint scaling of $k$ and $\rho$.
Unfortunately, the constant $c$ is enormous, scaling as $10^{450} E[S^3]$.
In contrast, we focus on the regime of fixed $k$,
and prove tight and explicit bounds on mean response time.
\citemanual{goldberg2017simple}
also provide a highly detailed literature review
on bounds on $E[T_Q]$ and related measures in the M/G/k and related models.

\subsection{Heterogeneous M/G/k}
\subsubsection{Heterogeneous M/M/k}
Much of the previous work on multiserver models with heterogeneous service rates
has focused on the much simpler M/M/k setting,
where jobs are memoryless
\citep{efrosinin_performance_2008,alves2011upper,efrosinin_approximations_2020,lin1984optimal}.
In this model,
one can analyze the preemptive Fastest-Server-First policy
to derive a natural lower bound on the mean response time of any server assignment policy.
One can similarly analyze the preemptive Slowest-Server-First policy
to derive an upper bound.
These two policies each lead to a single-dimensional birth-death Markov chain,
allowing for straightforward analysis \citep{alves2011upper}.
One can think of our bounds as essentially extending these bounds for the M/M/k
to the much more complex setting of the M/G/k.

\subsubsection{Heterogeneous M/H$_m$/k}
\citemanual{van2003markovian} primarily study a \emph{homogeneous} multiserver setting
with hyperexponential job sizes.
However, in their conclusion, they mention that their methods could be extended
to a setting with heterogeneous servers,
but at the cost of making their Markov chain grow exponentially.
This exponential blowup seems inevitable when applying exact Markovian methods to a heterogeneous setting
with differentiated jobs.

\subsubsection{M/(M+G)/2 Model}
Another intermediate model is the M/(M+G)/2 model of \citemanual{boxma2002waiting}.
In this model,
jobs are not differentiated.
Instead, the service time distribution is entirely dependent on the server.
Server 1, the first server to be used, has an exponential service time distribution,
while server 2 has a general service time distribution.
\citemanual{boxma2002waiting} derive an implicit expression for
the Laplace-Stieltjes transform of response time in this setting,
which they are only able to make explicit
when the general service time distribution has rational transform.
Subsequent work
has fully solved the M/(M+G)/2 model,
under both FCFS service and related service disciplines
\cite{keaogile_geo_2015,sani_mg2_2015,ramasamy_mg2_2015}.

Our results are not directly applicable to the M/(M+G)/2 setting,
because the servers have different distributions of service time, not just different speeds.
However, the slow progress on this two-server model
illustrates the immense difficulty in solving even the simplest
heterogeneous multiserver models.
In contrast, our WCFS framework handles both differentiated jobs
and an arbitrary number of servers with no additional effort.
\subsection{Limited Processor Sharing}
\label{sec:lps_prior}

The Limited Processor Sharing policy has been studied by a wide variety of authors
\cite{nuyens2008monotonicity,telek_response_2018,zhang_law_2009,zhang_diffusion_2011,zhang2008steady,gupta_2009_adaptive,harchol2013performance},
but none bound mean response time for all loads $\rho$.

\subsubsection{Asymptotic Regimes}
\label{sec:zhang}
A series of papers by Zhang, Dai and Zwart \cite{zhang_law_2009,zhang_diffusion_2011,zhang2008steady}
derive the strongest known results on Limited Processor Sharing in a variety of asymptotic regimes.
These authors derive the measure-valued fluid limit \cite{zhang_law_2009},
the diffusion limit \cite{zhang_diffusion_2011}
and a steady-state approximation \cite{zhang2008steady}.
The most comparable of their results to our work is their steady-state approximation.
When specialized to mean response time in the M/G/1/LPS,
their approximation states that
\begin{align*}
    E[T] &\approx \frac{E[S]}{1-\rho}(1-\rho^k)
    + \frac{E[S^2]}{2E[S]}\frac{\rho^k}{1-\rho}
\end{align*}

They prove that this approximation is accurate in the heavy-traffic limit;
they do not provide specific error bounds,
but empirically show the approximation performs well at all loads $\rho$ \cite{zhang2008steady}.
Our results therefore complement their results 
by proving concrete error bounds.

\subsubsection{State-dependent Server Speed}
To model the behavior of databases,
\citemanual{gupta_2009_adaptive} introduce a variant of the Limited Processor Sharing model,
where the total server speed is a function of the number of jobs in service.
In their setting server speed increases to a peak,
and then slowly declines as more jobs enter service.
They derive a two-moment approximation for mean response time,
and use it to derive a heuristic policy for choosing the Multi-Programming Level (MPL).
While this two-moment approximation is not known to be tight,
it indicates that the optimal MPL
for minimizing mean response time
may be significantly larger than the service-rate-maximizing MPL,
if job size variability is large and load is not too high.

Using our WCFS framework
it is possible to derive bounds on mean response time
for the state-dependent server speeds setting.
For MPL parameters less than or equal to the service-rate-maximizing MPL,
both our upper and lower bounds apply,
while if the MPL parameter is greater than the service-rate-maximizing MPL,
only our upper bounds apply,
because the system only partially fulfills our definition of work conservation.

Subsequently, \citemanual{telek_response_2018}
derive the Laplace-Stieltjes transform of response time
in the LPS model with state-dependent server speed,
under phase-type job sizes.
Unfortunately, the transform takes the form of a complicated matrix equation,
making it difficult
to derive general insights across general job size distributions.
Instead, the authors numerically invert the Laplace transform
for a handful of specific distributions to derive empirical insights.

\subsection{Threshold Parallelism}
\label{sec:parallelism_prior}
Jobs with ``speedup functions" are common in Machine Learning and other highly parallel computing settings.
A job's speedup function specifies the degree to which it can be parallelized. In 
\cite{berg_towards_2017, berg2021optimal, berg_optimal_2020}, the authors study optimal allocation policies of servers to jobs when the arriving jobs have different speedup function.  In many cases, a job's speedup function takes the form of a ``threshold" function: here the job receives perfect (linear) speedup up to some threshold number of servers and receives no additional speedup beyond that number of servers.  We refer to this as the Threshold Parallelism model.

While understanding the response time in systems where jobs have speedup functions is generally intractable, 
\citemanual{berg_optimal_2020} were able to approximately analyze response time in the case where every job is either
 ``inelastic,''
with parallelism threshold 1, or ``elastic,''
with parallelism threshold $k$.
They also assume that inelastic jobs have size distributed as $Exp(\mu_I)$,
and elastic jobs have size distributed as $Exp(\mu_E)$,
with sizes unknown to the scheduler.
They focus on two preemptive-priority service policies for this setting:
Inelastic First (IF) and Elastic First (EF).
In this setting, they approximate the mean response time 
of EF and IF within $1\%$ error by using a combination of the Busy-Period Transitions
technique and Matrix-Analytic methods
to evaluate their multidimensional Markov chain.

The Threshold Parallelism model in our paper is far broader than that in the prior literature, and our bounds are tighter in the heavy-traffic limit.  

\subsection{Multiserver Jobs}
\label{sec:multiserver_job_prior}
The Multiserver-Job model has been extensively studied,
in both practical \cite{feitelson_parallel_2004,srinivasan_characterization_2002,carastan_one_2019}
and theoretical settings \cite{brill1984queues,rumyantsev2017stability,hong2021sharp,maguluri2012stochastic,ghaderi2016randomized,psychas2018randomized,psychas2017non-preemptive,maguluri_scheduling_2014}.
It captures the common scenario in datacenters and supercomputing
where each job requires a fixed number of servers in order to run.
Characterizing the stability region of policies in this model is already a challenging problem,
and there were no bounds on mean response time for any scheduling policy,
prior to our bound on ServerFilling.

\subsubsection{FCFS Scheduling}
The most natural policy is FCFS,
where the oldest jobs are placed into service until a job requires more servers than remain,
at which point the queue is blocked.
Therefore, the FCFS policy can leave a large number of servers idle even when many jobs are present.
As a result, FCFS does not in general achieve an optimal stability region.
Even worse, deriving the stability region of FCFS is an open problem,
and has only been solved in a few special cases \cite{brill1984queues,rumyantsev2017stability}.

One technique that may be useful for characterizing this stability region
is the \textit{saturated system} approach \cite{baccelli_foss_1995,foss2004overview}.
The saturated system is a system in which additional jobs are always available,
so the front is always full,
only the composition of jobs in the front varies.
The completion rate of the saturated system exactly matches
the stability region of the equivalent open system, under a wide variety of arrival processes.
Unfortunately, analyzing the general Multiserver-Job FCFS saturated system seems intractable.

Given the difficulty of proving results under FCFS scheduling,
finding policies with better theoretical guarantees, such as ServerFilling, is desirable.

\subsubsection{MaxWeight Scheduling}
\label{sec:max-weight}
One natural throughput-optimal policy is the MaxWeight policy \cite{maguluri2012stochastic}.
Here jobs are divided into classes based on their server requirements,
with $N_i(t)$ denoting the number of jobs requiring $i$ servers in the system at time $t$.  Let the set $Z(t)$ denote all possible packings of jobs at time $t$ onto servers.  Let $z \in Z(t)$ be a particular packing, where  $z_i$ denotes the number of jobs requiring $i$ servers that are 
served by packing $z$.

The MaxWeight service policy picks the packing $z$ which maximizes
\begin{align*}
    \max_z \sum_i N_i(t) z_i.
\end{align*}

For example, if there are many jobs requiring $3$ servers, we want to pick a packing that serves many $3$-server jobs. 
While MaxWeight is throughput optimal,
it is very computationally intensive to implement,
requiring the scheduler to solve
an NP-hard optimization problem whenever a job arrives or departs.
For comparison, ServerFilling is also throughput-optimal given our 
assumptions on the server requirements $V$,
but it is far computationally simpler, requiring approximately linear time as a function of $k$.
Moreover, no bounds on mean response time are known for MaxWeight,
due in part to its high complexity.

\subsubsection{Nonpreemptive Scheduling}
In certain practical settings such as supercomputing,
a nonpreemptive service policy is preferred.
In such settings, a backfilling policy such as EASY backfilling or conservative backfilling is often used
\cite{feitelson_parallel_2004,srinivasan_characterization_2002,carastan_one_2019}.
These start by serving jobs in FCFS order,
until a job is reached that requires more servers than remain.
At this point, jobs further back in the queue that require fewer servers are scheduled,
but only if they will not delay older jobs,
based on user-provided service time upper bounds.
While these policies are popular in practice
little is known about them theoretically, including their response time characteristics.

Finding any nonpreemptive throughput-optimal policy is a challenging problem.
Several such policies have been designed \cite{maguluri_scheduling_2014,ghaderi2016randomized,psychas2018randomized},
typically by slowly shifting between different server configurations to alleviate overhead.
Because such policies can have very large renewal times,
many jobs can back up while the system is in a low-efficiency configuration,
which can empirically lead to very high mean response times.
However, no theoretical mean response time analysis
exists for any policy in the Multiserver-Job setting.
As a result, there is no good baseline policy to compare against novel policies,
Our bounds on the mean response time of ServerFilling
can serve as such a baseline,
albeit in the more permissive setting of preemptive scheduling.

\section{Theorems and Proofs}
\label{sec:results}
We perform a heavy traffic analysis within our WCFS framework, assuming
finite $\rem_{\sup}(S, C)$.
Specifically,
we prove that the scaled mean response time of any WCFS
model converges to the same constant as an M/G/1/FCFS:
\begin{theorem}[Heavy Traffic response time]
\label{thm:scaled-et}
    For any model $\pi \in$ WCFS, if $\rem_{\sup}(S, C)$ is finite,
\begin{align*}
    \lim_{\rho \to 1} E[T^\pi] (1-\rho) = \frac{E[S^2]}{2E[S]}.
\end{align*}
\end{theorem}

To prove \cref{thm:scaled-et},
we prove a stronger theorem,
tightly and explicitly bounding $E[T^{\pi}]$
up to an additive constant, for any $\pi \in$ WCFS.

\begin{theorem}[Explicit response time bounds]
    \label{thm:explicit-bounds}
    For any model $\pi \in$ WCFS, if $\rem_{\sup}(S, C)$ is finite,
    \begin{align*}
    E[T^\pi] &\le\frac{\rho}{1-\rho}\frac{E[S^2]}{2 E[S]} + c_{upper}^\pi \\
    E[T^\pi] &\ge\frac{\rho}{1-\rho}\frac{E[S^2]}{2 E[S]} + c_{lower}^\pi
    \end{align*}
for explicit constants $c_{upper}^\pi$ and $c_{lower}^\pi$ not dependent on load $\rho$.
\end{theorem}
\begin{proof}[Proof deferred to \cref{sec:proof-of-explicit-bounds}]
\end{proof}

From \cref{thm:explicit-bounds}, \cref{thm:scaled-et} follows
via a simple rearrangement:
\begin{align*}
     \frac{\rho}{1-\rho} \frac{E[S^2]}{2E[S]}
    = \frac{\frac{E[S^2]}{2E[S]}}{1-\rho} - \frac{E[S^2]}{2E[S]}.
\end{align*}

\cref{thm:explicit-bounds}
also implies rapid convergence of scaled mean response time
to its limiting constant for any WCFS policy:
\begin{corollary}
\label{cor:rapid-convergence}
    For any model $\pi \in$ WCFS, if $\rem_{\sup}(S, C)$ is finite,
\begin{align*}
    E[T^\pi] (1-\rho) = \frac{E[S^2]}{2E[S]} + O(1-\rho).
\end{align*}
\end{corollary}
\subsection{Outline of Proof of \cref{thm:explicit-bounds}}
\label{sec:proof-of-explicit-bounds}
We will prove \cref{thm:explicit-bounds}
where
\begin{align*}
    c_{upper}^\pi &= (n-1)\rem_{\sup}(S, C) + \frac{n E[S]}{b_{\inf}}, \\
    c_{lower}^\pi &= - (n-1)\rem_{\sup}(S, C) + E[S],
\end{align*}
where $n$ denotes the size of the front,
and where $b_{\inf}$ is defined in \cref{sec:positive-b-min}.

Our goal is simply to prove the bounds in \cref{thm:explicit-bounds}
for some constants $c^\pi_{upper}, c^\pi_{lower}$
independent of $\rho$;
we have made no effort to optimize these constants, leaving that to future work.
Specifically, for three of our four motivating models,
the $\frac{n}{b_{\inf}}$ term scales as $O(n^2)$.
For these models this term is unnecessarily loose,
and could easily be lowered to an $O(n)$ bound by using a more detailed view.

Our approach is to split response time $T$ into two pieces,
queueing time $T_Q$ and front time $T_F$,
and bound the expectation of each separately.
We first bound $E[T_Q]$,
which forms the bulk of our proof.
The two key ideas
come from the intuition that a WCFS model behaves like a FCFS M/G/1 system.
In \cref{lem:queueing-work}, we prove that 
$E[T_Q] = E[W] + c$, for some constant $c$;
in a WCFS model,
jobs progress through the system in essentially FCFS order,
and as $\rho \to 1$ work is completed essentially at rate 1.

In \cref{lem:work-bounds}, we prove that
$E[W] = E[W^{M/G/1}] + c$, for some constant $c$.
The key idea here is that in a WCFS model,
if $W$ is large,
work arrives and completes in exactly
the same way as in an M/G/1.
Likewise,
if the front is not full, then $W$ cannot be large.

In \cref{lem:queueing-bounds}, we combine \cref{lem:queueing-work,lem:work-bounds}
to prove that $E[T_Q] = E[T^{M/G/1}] + c$ for some constant $c$.

In \cref{lem:finite-work},
we prove that work $W$ is indeed stationary with finite mean.
This is a technical lemma that rules out pathological scenarios,
which is necessary because our WCFS class of models is very general.
\cref{lem:finite-work} is used by both \cref{lem:queueing-work,lem:work-bounds}.

Finally, in \cref{lem:front-bounds},
we bound $E[T_F]$, utilizing Little's law.

Combining \cref{lem:queueing-bounds,lem:front-bounds} proves \cref{thm:explicit-bounds}.

\subsection{Two Views}
\label{sec:two-views}

At several steps in our proof of \cref{thm:explicit-bounds},
we will make use of two different views of the queueing system,
corresponding to two different state descriptors:
\begin{description}
    \item[Omniscient view:] In the omniscient view
    the state descriptor consists
    of the remaining size and class of all jobs in the system;
    we sample jobs' sizes and classes when the jobs enter the system.
    For a given system state, work is a deterministic quantity.
    \item[Limited view:] In the limited view,
    the state descriptor consists
    of the age and class of the jobs in the front,
    and the number of jobs in the queue.
    We sample jobs' classes when they enter the front,
    and determine whether jobs complete according
    to the hazard rate of the job size distribution,
    as the job ages.
    For a given system state, work is a random variable.
\end{description}
We will make it clear which view of the system we are using in each step of the proof.
Generally, the omniscient view is useful when analyzing
total work in the system,
and the limited view is useful when
analyzing work at the front.

\subsection{\cref{lem:queueing-work}: $E[T_Q]$ and $E[W]$}
First, we prove that mean queueing time and mean work are similar:

\begin{lemma}[Queueing time and work]
    \label{lem:queueing-work}
    For any model $\pi \in$ WCFS, if $\rem_{\sup}(S, C)$ is finite,
    \begin{align*}
        E[W] - (n-1)\rem_{\sup}(S, C) \le E[T_Q] \le E[W]. 
    \end{align*}
\end{lemma}
\begin{proof}
    Start by writing time in queue $T_Q$ in terms of work in system.
    Let us consider the omniscient view of the system,
    so work $W$ is a deterministic quantity given the system state.
    Consider an arbitrary tagged job $j$.
    When $j$ arrives,
    let $W^A(j)$ be the amount of work $j$ sees in the system.
    Let $W^F_F(j)$ be the amount of work $j$ sees in the front other than $j$ itself,
    when $j$ leaves the queue and enters the front.
    In $W^F_F$, the subscript $F$ indicates that we are looking at the amount of work at the front,
    and the superscript $F$ indicates that we are looking at the moment when $j$
    enters the front.

    Because the model is finite-skip,
    jobs move from the queue to the front in arrival order,
    so all of the $W^A(j)$ work that was in the system when $j$ arrived is either
    complete or in the front when $j$ enters the front.
    As a result, the amount of work which is completed while $j$ is in the queue
    is exactly $W^A(j) - W^F_F(j)$.
    Note that if $j$ enters the front upon arrival to the system,
    $W^A(j) = W^F_F(j)$,
    and no work is completed while $j$ is in the queue.

    While $j$ is in the queue, the front must be full;
    the system must be maximally busy during this time,
    completing work at rate 1.
    Job $j$ is in the queue for $T_Q(j)$ time,
    so the system must complete $T_Q(j)$ work during that time.
    We can therefore conclude that
    \begin{align*}
        W^A(j) - W^F_F(j) = T_Q(j).
    \end{align*}
    Because $j$ is an arbitrary job, we can write $W^F_F(j)$ as $W^F_F$,
    a random variable over all jobs that pass through the system.
    Likewise, $T_Q(j)$ is simply $T_Q$.
    Because Poisson arrivals see time averages,
    $W^A(j) \sim W$, the time-stationary amount of work in the system.
    Combining these equivalencies,
    we find that
    \begin{align}
        \label{eq:work-to-time}
        W - W^F_F = T_Q.
    \end{align}
    Note that $W$ is time-stationary,
    while $W_F^F$ and $T_Q$ are event-stationary.

    To rigorously demonstrate \eqref{eq:work-to-time},
    we need to prove that the system converges to a stationary distribution,
    which we prove in \cref{lem:finite-work}.

    To give bounds on $W^F_F$,
    we switch to the limited view of the system,
    where the state of the front consists of the classes and ages of the jobs at the front.
    We have two simple bounds on $W^F_F$:
    First, $W^F_F \ge 0$.
    Next, because $W^F_F(j)$ is the work of at most $n-1$ jobs,
    the jobs at the front when a given job enters the front,
    we know that
    \begin{align*}
        E[W^F_F] \le (n-1)\rem_{\sup}(S, C).
    \end{align*}

    Combining these bounds with \eqref{eq:work-to-time},
    we can bound $E[T_Q]$ in terms of $E[W]$:
    \begin{align*}
        E[W] - (n-1)\rem_{\sup}(S, C) \le E[T_Q] \le E[W]. 
    \end{align*}
\end{proof}

\subsection{\cref{lem:work-bounds}: Bounding $E[W]$}
\begin{lemma}(Work bounds)
    \label{lem:work-bounds}
    For any model $\pi \in$ WCFS, if $\rem_{\sup}(S, C)$ is finite,
    \begin{align*}
        \frac{\rho}{1-\rho} \frac{E[S^2]}{2E[S]} \le E[W] \le \frac{\rho}{1-\rho} \frac{E[S^2]}{2E[S]}
        + (n-1) \rem_{\sup}(S, C).
    \end{align*}
\end{lemma}
\begin{proof}
    Consider the stationary random variable $W^2$
    in the omniscient view,
    so work is a deterministic quantity at a given time on a given sample path.
    $W^2$ evolves in two ways: continuous decrease as work is completed,
    and stochastic jumps as jobs arrive.
    Because $W^2$ is a stationary random variable,
    the expected rate of decrease and increase must be equal,
    due to the rate conservation law \cite{miyazawa1994rate}
    with respect to $W^2$.

    To calculate the expected rate of decrease,
    note that, ignoring moments where jobs arrive,
    $\frac{d}{dt} W(t) = -B(t)$,
    by definition, where
    $B(t)$ is the total service rate of the system at time $t$.
    As a result,
    $\frac{d}{dt} W(t)^2 = -2 W(t) B(t)$,
    ignoring arrival epochs.
    This expected rate of decrease is a well-defined random variable,
    because the system converges to stationarity.
    Thus the expected rate of decrease of $W^2$
    is $2E[WB]$.

    To calculate the expected rate of increase,
    let $t^-$ be the time just before a job arrives to the system.
    When the job arrives, $W^2$ increases from $W(t^-)^2$
    to $(W(t^-) + S)^2$,
    a change of $2W(t^-)S + S^2$.
    Note that $W(t^-)$ is distributed as $W$,
    by PASTA.
    Note also that $W$ and $S$ are independent,
    because $S$ is sampled i.i.d..
    As a result, the expected increase per arrival is
    $2E[W]E[S] + E[S^2]$.
    Arrivals occur at rate $\lambda$.
    As a result, the expected rate of increase
    is $2\lambda E[W]E[S] + \lambda E[S^2]$.

    To show that these rates are equal,
    we must show that the rates are finite.
    This follows from the fact that $E[W]$ is finite,
    which we prove in \cref{lem:finite-work}.

    As a result, the rates of increase and decrease of $W^2$ are equal:
    \begin{align}
    \nonumber
        2E[WB] &= 2\lambda E[W]E[S] + \lambda E[S^2] \\
    \nonumber
        E[WB] &= \lambda E[W]E[S] + \frac{\lambda}{2} E[S^2] \\
    \nonumber
        E[WB] &= \rho E[W] + \frac{\lambda}{2} E[S^2] \\
    \nonumber
        E[W] - E[W(1-B)] &= \rho E[W] + \frac{\lambda}{2} E[S^2] \\
    \nonumber
        E[W](1-\rho) &= E[W(1-B)] + \frac{\lambda}{2} E[S^2] \\
    \label{eq:mean_work}
        E[W] &= \frac{E[W(1-B)]}{1-\rho} + \frac{\lambda E[S^2]}{2(1-\rho)}
    \end{align}

    Now, we merely need to bound $E[W(1-B)]$.
    We do so by switching to the limited view.
    Note that
    \begin{align*}
        E[W(1-B)] &= E[W(1-B) \mathbbm{1}\{B=1\}] + E[W(1-B) \mathbbm{1}\{B < 1\}] \\
        &= E[W(1-B) \mathbbm{1}\{B < 1\}]
    \end{align*}

    Because the model is work-conserving,
    if $B < 1$, the front is not full,
    and there are at most $n-1$ jobs in the system.
    Taking expectations over the future randomness of these jobs,
    at any time $t$
    for which $B(t) < 1$,
    \begin{align*}
        E[W(t)] \le (n-1)\rem_{\sup}(S, C)
    \end{align*}
    Therefore,
    \begin{align*}
        E[W(1-B)\mathbbm{1}\{B < 1\}] &\le (n-1)\rem_{\sup}(S, C) E[(1-B)\mathbbm{1}\{B < 1\}] \\
        &= (n-1)\rem_{\sup}(S, C) E[1-B] \\
        &= (n-1)\rem_{\sup}(S, C) (1-\rho) \\
        E[W(1-B)] &\le (n-1)\rem_{\sup}(S, C) (1-\rho).
    \end{align*}
    Substituting this into \eqref{eq:mean_work}, our equation for $E[W]$,
    we find that
    \begin{align*}
        E[W] &\le \frac{\lambda E[S^2]}{2(1-\rho)} + (n-1)\rem_{\sup}(S, C).
    \end{align*}
    Dropping the first term of \eqref{eq:mean_work},
    we also get a lower bound:
    \begin{align*}
        E[W] &\ge \frac{\lambda E[S^2]}{2(1-\rho)}.
    \end{align*}
\end{proof}

One might alternatively try to prove \cref{lem:work-bounds}
via a coupling argument, by coupling the WCFS system to an M/G/1 with the same arrival process.
Unfortunately, this proof strategy does not succeed, for a subtle reason.

One can show that the difference in work between the two systems
during an interval when the WCFS system has a full front
is bounded by the amount of work in the WCFS system
at the beginning of the interval.
This is analogous to the many-jobs interval argument used by Grosof et al. \cite{grosof2018srpt}
to analyze relevant work in the M/G/k/SRPT.
The key difference is that in the WCFS setting, we consider total work, not relevant work,
meaning that job sizes are not bounded.
As a result, while the expected work at the beginning of a full-front interval is bounded,
the realization of that work may be arbitrarily large.

A coupling argument would therefore need to bound the relative
length of full-front intervals started by
different amounts of work,
to prove a time-average bound on the gap between $E[W]$ and $E[W^{M/G/1}]$.
This seems intractable, given the generality of WCFS policies.

Instead, by using a rate-conservation approach, formalized by Palm Calculus,
we directly connect
the small expected amount of work in a WCFS system with non-full front
to a small expected difference in work between the two systems.
We therefore prove \cref{lem:work-bounds},
while avoiding all of the complications of a coupling-based argument.

\subsection{\cref{lem:queueing-bounds}: Bounding $E[T_Q]$}

Now, we can bound $E[T_Q]$ by combining \cref{lem:queueing-work,lem:work-bounds}:
\begin{lemma}[Queueing time bounds]
    \label{lem:queueing-bounds}
    For any model $\pi \in$ WCFS, if $\rem_{\sup}(S, C)$ is finite,
    \begin{align*}
        E[T^\pi_Q] &\le \frac{\rho}{1-\rho} \frac{E[S^2]}{2E[S]} + (n-1)\rem_{\sup}(S, C) \\
        E[T^\pi_Q] &\ge \frac{\rho}{1-\rho} \frac{E[S^2]}{2E[S]} - (n-1)\rem_{\sup}(S, C)
    \end{align*}
\end{lemma}
\subsection{\cref{lem:finite-work}: Finite $E[W]$}
\begin{lemma}[Finite mean work]
\label{lem:finite-work}
    For any model $\pi \in$ WCFS, if $\rem_{\sup}(S, C)$ is finite,
    for any load $\rho < 1$,
    $W$ is a well-defined stationary random variable and
    $E[W]$ is finite.
\end{lemma}
\begin{proof}
Recall that $W = W_F + W_Q$;
we first focus on $W_F$.
There are at most $n$ jobs in the front at any time.
In the limited view,
each job has expected remaining size at most $\rem_{\sup}(S, C)$,
so $E[W_F] \le n \rem_{\sup}(S, C)$.

As for the stationarity of the state of the front,
this follows from two assumptions we made in \cref{sec:finite-rem-sup}.
First, we assumed that the service policy
is dependent only on the state of the front.
Second, the front must empty and thereby undergo renewals,
because the service rate $B(t)$ is at least $b_{\inf}$ whenever the system is nonempty.
As a result, $W_F$ is stationary.

We now turn to $W_Q$.
To prove that $W_Q$ is stationary and well-defined with finite mean,
we will apply the ``inventory process'' results of \citemanual{sigman1994finite},
and \citemanual{scheller_finite_1996}'s refinement of those results.

We upper bound $W_Q$ by $\cW$, which we will write as an inventory process.
\begin{align*}
    \cW := W \mathbbm{1}\{W_Q > 0\}.
\end{align*}
Here we will use the omniscient view, so $\cW(t)$ is a specific value.
By proving $\cW$ is stationary and well-defined with finite mean,
we also show the same is true of $W_Q$.
Because $W_Q = (\cW - W_F)^+$, the stationarity of $\cW$
also implies the stationarity of $W_Q$, given the stationarity of $W_F$.

To write $\cW$ as an inventory process as in \cite{sigman1994finite},
we must define a process $X(t)$ with stationary and ergodic increments,
such that
\begin{align*}
    \cW(t) = X(t) + L(t),
\end{align*}
where
\begin{align*}
    L(t) := \sup_{0 \le s \le t} (-\min \{0, X(s)\}).
\end{align*}

Here $X(t)$ represents the potential workload process,
and $L(t)$ corrects for the fact that the queue can empty.

We will apply \cite[Theorem 2.2.1]{scheller_finite_1996},
for the special case of the first moment.
Note by Remarks 1 and 3,
for the first moment of an inventory process,
it suffices to show:
\begin{itemize}
    \item Negative drift: There exists an amount of work $w < \infty$
    and a drift rate $\delta > 0$ such that
    conditioned on $\cW(t) \ge w$,
    \begin{align*}
        \lim_{\epsilon \to 0} \frac{E_{\cF_t}[X(t+\epsilon)-X(t)]}{\epsilon} \ge - \delta
    \end{align*}
    where $\cF_t$ is the filtration defined by the behavior of the system up to time $t$.
    \item Finite second moment of positive jumps:
        There exists a constant $k_1 < \infty$ such that 
        \begin{align*}
            \lim_{\epsilon \to 0} E_{\cF_t}[((X(t+\epsilon)-X(t))^+)^2] \le k_1
        \end{align*}
\end{itemize}

Now, we define the potential workload process $X(t)$
based on $W(t)$ and $W_Q(t)$.

During intervals when $W_Q(t) = 0$, $X(t)$ is constant.
If $t_0$ is the beginning of an interval where $W_Q(t) > 0$,
$X(t)$ jumps up by $W(t_0^+)$ at time $t_0$.
During an interval where $W_Q(t) > 0$,
$X(t)$ mimics $W(t)$:
$X(t)$ rises by $S$ when a job arrives,
and decreases at rate 1.
If $t_1$ is the end of an interval where $W_Q(t) > 0$,
$X(t)$ jumps down by $W(t_1^-)$ at time $t_1$.

By construction, $X(t)$ generates $\cW(t)$ as an inventory process.
For example, let $t_1$ be the end of an interval where $W_Q(t) > 0$.
Assume that the desired relationship between $X(t)$ and $\cW(t)$
holds up to time $t_1^-$. In particular, $\cW(t_1^-) = W(t_1^-)$.
Then $\cW(t_1^+) = 0$, as desired.

Next, we show that $X(t)$ has stationary and ergodic increments.
$X(t)$ has two types of increments:
First, Poisson arrivals cause increments sampled i.i.d. from $S$,
which are clearly stationary and ergodic.
Second, the beginning and end of intervals where $W_Q(t) = 0$
cause increments equal to $W_F(t)$.
These increments are stationary and ergodic
because the state of the front, and $W_F$ in particular, are stationary.
Thus, $X(t)$ has stationary and ergodic increments.

To demonstrate negative drift, let $w$ be an arbitrary nonzero amount of work.
Whenever $\cW(t) \ge w$,
$X(t)$ has two types of increments: jumps of size $S$ occurring at rate $\lambda$,
and continuous decrease at rate 1.
As a result, the drift of $X(t)$ is $\rho - 1 < 0$.

To demonstrate finite second moment of positive jumps,
note that $X(t)$ has two kinds of positive jumps:
Jumps of size $S$, when $W_Q(t) > 0$,
and jumps of size $W(t)$, at the beginning of a $W_Q > 0$ interval.

Switching back to the limited view, note that
the latter kind of jump consists of the remaining size of at most $n$ jobs.
These remaining sizes are distributed as
\begin{align*}
    R(a, c) \sim [S_c - a \mid S_c > a]
\end{align*}
for some age $a$ and class $c$.

It therefore suffices to show that there exists a constant $r$ such that for all $a, c$,
\begin{align*}
    E[R(a, c)^2] \le r < \infty.
\end{align*}

To do so, we will write $R(a, c)_e$,
the excess of the remaining size distribution,
as a mixture of remaining size distributions for different ages.
Note that for any distribution $Y$,
the excess $Y_e$ is equivalent to
\begin{align*}
    Y_e \sim [Y - Y_e \mid Y > Y_e].
\end{align*}
This holds because the forward and backwards renewal times are distributed identically \cite[Chapter 23]{harchol2013performance}.
By applying this construction with $Y=R(a, c)$, we find that
\begin{align*}
    R(a, c)_e &\sim [R(a, c) - R(a, c)_e \mid R(a, c) > R(a, c)_e] \\
    &= [S_c - (a + R(a, c)_e) \mid S_c > a + R(a, c)_e].
\end{align*}
As a result, $a + R(a, c)_e$ is the desired age distribution.

For any age $a'$, $E[R(a', c)] \le \rem_{\sup}(S, C)$.
Because $R(a, c)_e$ can be written as a mixture of remaining size distributions,
$E[R(a, c)_e] \le \rem_{\sup}(S, C)$,
which is finite by assumption.

We can now bound $E[R(a, c)^2]$:
\begin{align*}
    E[R(a,c)_e] &= \frac{E[R(a,c)^2)]}{2E[R(a,c)]} \\
    E[R(a, c)^2] &= 2E[R(a, c)]E[R(a,c)_e] \le 2 \rem_{\sup}(S, C)^2
\end{align*}

Thus, the requirements of \cite[Theorem 2.2.1]{scheller_finite_1996} are satisfied,
so both $\cW$ and $W_Q$ are stationary and well-defined, and have finite mean.

\end{proof}

\subsection{\cref{lem:front-bounds}: Bounding $E[T_F]$}
\begin{lemma}[Front time bounds]
    \label{lem:front-bounds}
    For any model $\pi \in$ WCFS,
    \begin{align*}
        E[S] \le E[T^F] &\le \frac{n E[S]}{b_{\inf}}
    \end{align*}
\end{lemma}
\begin{proof}
    First, to prove that $E[T^F] \ge E[S]$,
    note that if a job receives service at the maximum possible rate of 1
    for the entire time it is in the front,
    then the job will complete in time $S$.
    As a result, $E[T^F] \ge E[S]$.

    To prove the upper bound,
    recall that by the non-idling
    assumption from \cref{sec:positive-b-min},
    in all states of the front $s$
    where $N_F(s) \ge 1$,
    the service rate $B(s) \ge b_{\inf}$.
    Because $N_F(s) \le n$,
    we can bound the ratio $B(s)/N_F(s)$ in all $N_F(s) \ge 1$ states:
    \begin{align*}
        \frac{B(s)}{N_F(s)} &\ge \frac{b_{\inf}}{n}.
    \end{align*}

    Therefore, in all states, 
    \begin{align*}
        B(s) &\ge \frac{b_{\inf}}{n} N_F(s).
    \end{align*}
    In expectation,
    the same must hold:
    \begin{align*}
        E[B] \ge \frac{b_{\inf}}{n} E[N_F].
    \end{align*}
    Note that $E[B] = \rho$ and $E[N_F] = \lambda E[T_F]$
    by Little's Law.
    Thus,
    \begin{align*}
        \rho &\ge \frac{b_{\inf}}{n} \lambda E[T_F] \\
        \frac{n E[S]}{b_{\inf}} &\ge E[T_F].
    \end{align*}
\end{proof}
Note that \cref{lem:front-bounds}
proves a relatively weak bound on $E[T^F]$,
because we have only made the weak assumption that $b_{\inf}$ is positive.
In many models,
one can prove a stronger bound on $E[T^F]$
by using more information about the model's dynamics
when the front is not full.

From \cref{lem:queueing-bounds} and \cref{lem:front-bounds},
\cref{thm:explicit-bounds} follows immediately,
with explicit formulas for $c^\pi_{upper}$ and $c^\pi_{lower}$.

\section{Empirical Comparison: WCFS and non-WCFS}
\label{sec:empirical}
\begin{figure}[t]
    \centering
    \includegraphics[width=0.7\textwidth]{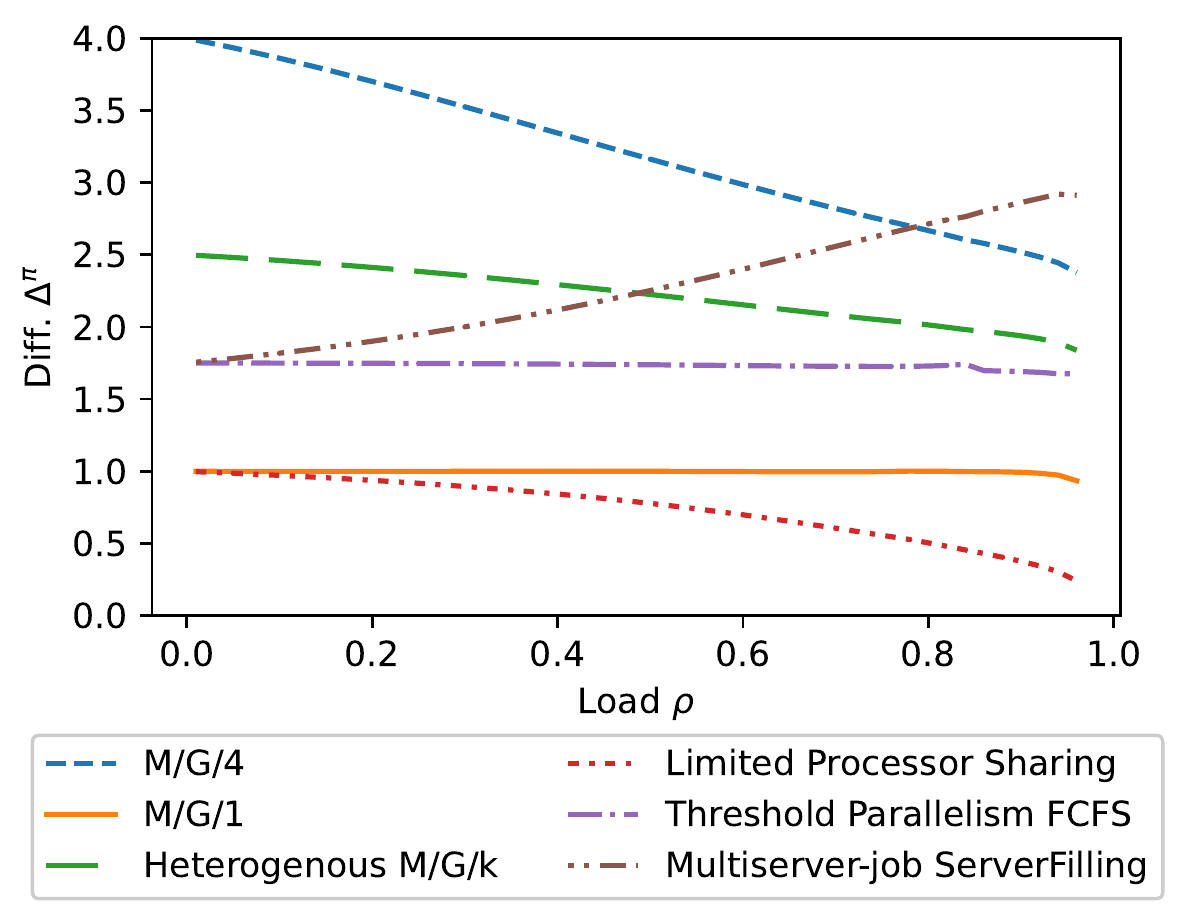}
    \caption{$\Delta^\pi$ for WCFS models.
Job size distribution $S$ is hyperexponential:
$Exp(2)$ w.p. $1/2$, $Exp(2/3)$ otherwise.
$10^9$ arrivals simulated.
$\rho > 0.96$ omitted due to the large amount of random noise under high load.
    Specific settings:
    Heterogeneous M/G/k with speeds $[0.4, 0.3, 0.2, 0.1]$.
    Limited Processor Sharing with Multi-programming Level 4.
    Threshold Parallelism FCFS with joint random variable $(S, L)$
    of $(Exp(2), 1)$ w.p. 1/2, $(Exp(2/3), 4)$ otherwise.
    Multiserver-job ServerFilling with joint random variable $(V, X)$
    of $(1, Exp(1/2))$ w.p. 1/2, $(4, Exp(2/3))$ otherwise.
}
    \label{fig:diff-wcfs}
\end{figure}

We have proven tight bounds on mean response time for all WCFS policies.
To quantify the tightness of our bounds,
we define the \emph{mean response time difference} $\Delta^\pi$
for a given policy $\pi$:
\begin{align*}
    \Delta^\pi = E[T^\pi] - \frac{\rho}{1-\rho}\frac{E[S^2]}{2 E[S]} = E[T^\pi] - E[T^{M/G/1}_Q].
\end{align*}
For instance, $\Delta^{M/G/1} = E[S]$.

This definition is useful because we have shown in \cref{thm:explicit-bounds}
that for any load $\rho$, $\Delta^\pi \in [c^\pi_{lower}, c^\pi_{upper}]$,
for constants $c^\pi_{lower}, c^\pi_{upper}$ not dependent on $\rho$,
but potentially depending on the model $\pi$.

To investigate the behavior of $\Delta^\pi$,
we turn to simulation.
We simulate both WCFS models, to confirm our results,
as well as non-WCFS models, to show that non-WCFS models typically do not
have constant $\Delta^\pi$ in the $\rho \to 1$ limit.

In \cref{fig:diff-wcfs},
we simulate WCFS models:
our four motivating models from \cref{sec:special-models},
as well as the simpler M/G/k and M/G/1 models.
In each case, we find that $\Delta^\pi$
remains bounded quite close to $0$,
meaning that \cref{thm:explicit-bounds}
holds with constants close to 0.

In \cref{fig:diff-wcfs},
we see that for some models, $\Delta^\pi$ increases with $\rho$,
while for others, $\Delta^\pi$ decreases with $\rho$.
Intuitively, this depends on which jobs tend to be prioritized as $\rho \to 1$.
Policies which serve many jobs at once, such as the $M/G/4$
and Limited Processor Sharing systems,
typically have $\Delta^\pi$ decrease as $\rho \to 1$,
because they allow small and large jobs to share service.
As a result, small jobs can complete faster than in an M/G/1,
lowering $\Delta^\pi$ if $\rho$ is large enough that
many jobs are typically in the system.

In contrast, policies which reorder large jobs ahead of small jobs
typically have $\Delta^\pi$ increase as $\rho \to 1$,
by the same principle.
For example, Multiserver-Job ServerFilling
prioritizes jobs in the front which require 4 servers.
In the setting depicted in \cref{fig:diff-wcfs},
such jobs have mean size $3/2$ in this system,
compared to the overall mean size $E[S] = 1$.

\begin{figure}
\begin{subfigure}{0.5\textwidth}
    \centering
    \includegraphics[width=\linewidth]{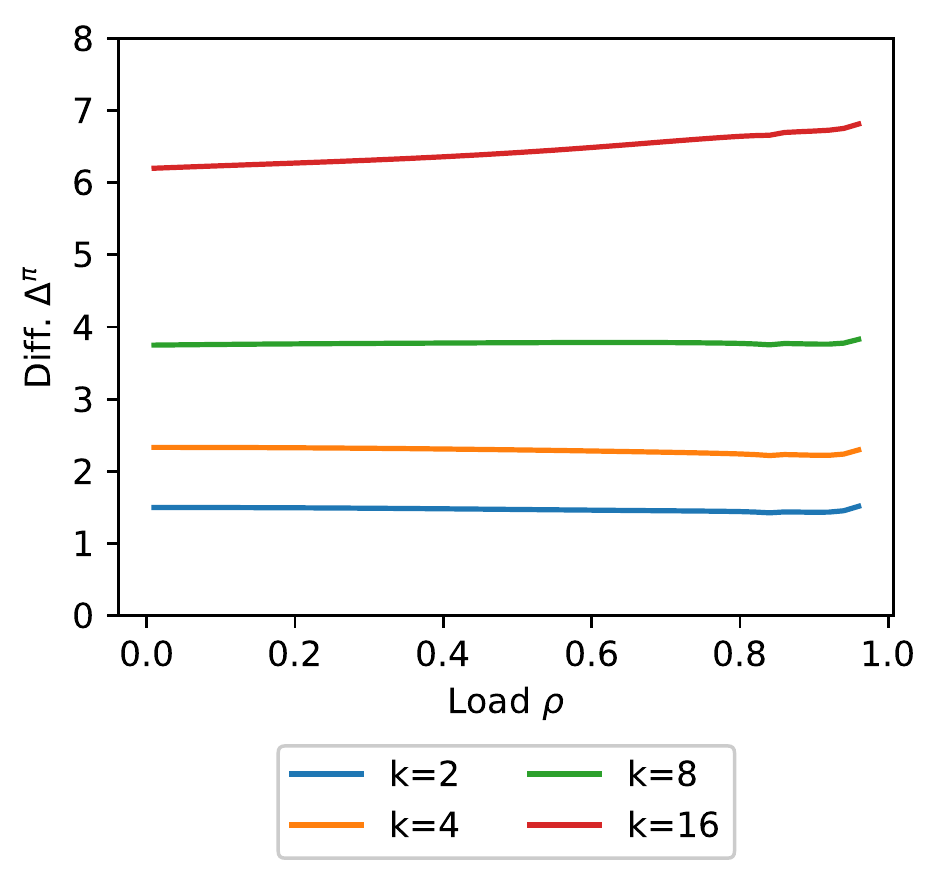}
    \caption{Varying front size $n$.
    Multiserver-job ServerFilling
    with $k=[2, 4, 8, 16]$.
    $S$ distributed $Exp(1)$.
    Server requirement $V$ distributed uniformly over all integer powers of 2 $\le k$.
    }
    \label{fig:many-k}
\end{subfigure}\quad
\begin{subfigure}{0.5\textwidth}
    \centering
    \includegraphics[width=\linewidth]{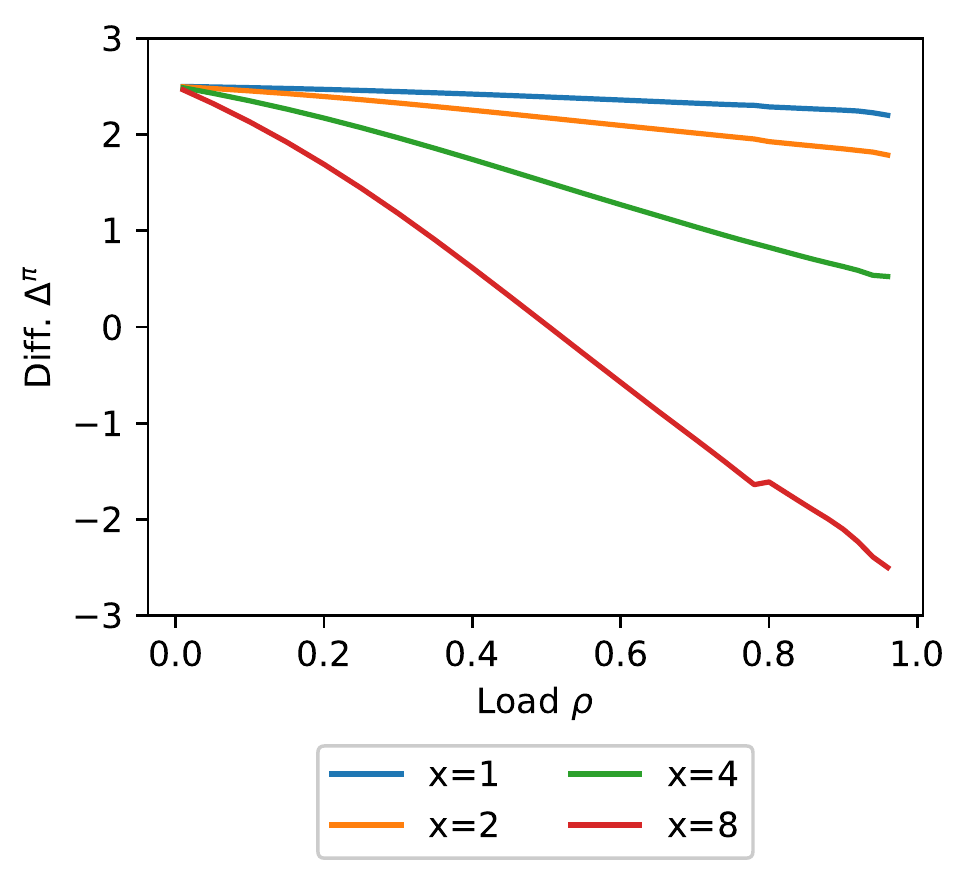}
    \caption{Varying job size distributions.
    Heterogeneous M/G/4
    with speeds $[0.4, 0.3, 0.2, 0.1]$.
    $S$ distributed
    hyperexponential: $Exp(1/x)$ with probability $1/2x$,
    else $Exp((2x-1)/x)$,
    for $x \in [1, 2, 4, 8]$.
    $E[S]=1, C^2 \approxeq [1, 1.67, 3.57, 7.53]$.
    }
    \label{fig:many-s}
\end{subfigure}
\caption{$\Delta^\pi$ under WCFS models with varying conditions.
Up to $10^9$ arrivals simulated.}
\end{figure}

In all of the settings simulated in \cref{fig:diff-wcfs}, $\Delta^\pi > 0$.
This is merely a coincidence,
not a general rule, as can be seen in \cref{fig:many-s}.

Regardless of the different reordering behavior of these different WCFS policies,
$\Delta^\pi$ does not diverge as $\rho \to 1$, as predicted by \cref{thm:explicit-bounds}.

\begin{figure}[t]
    \centering
    \includegraphics[width=0.9\textwidth]{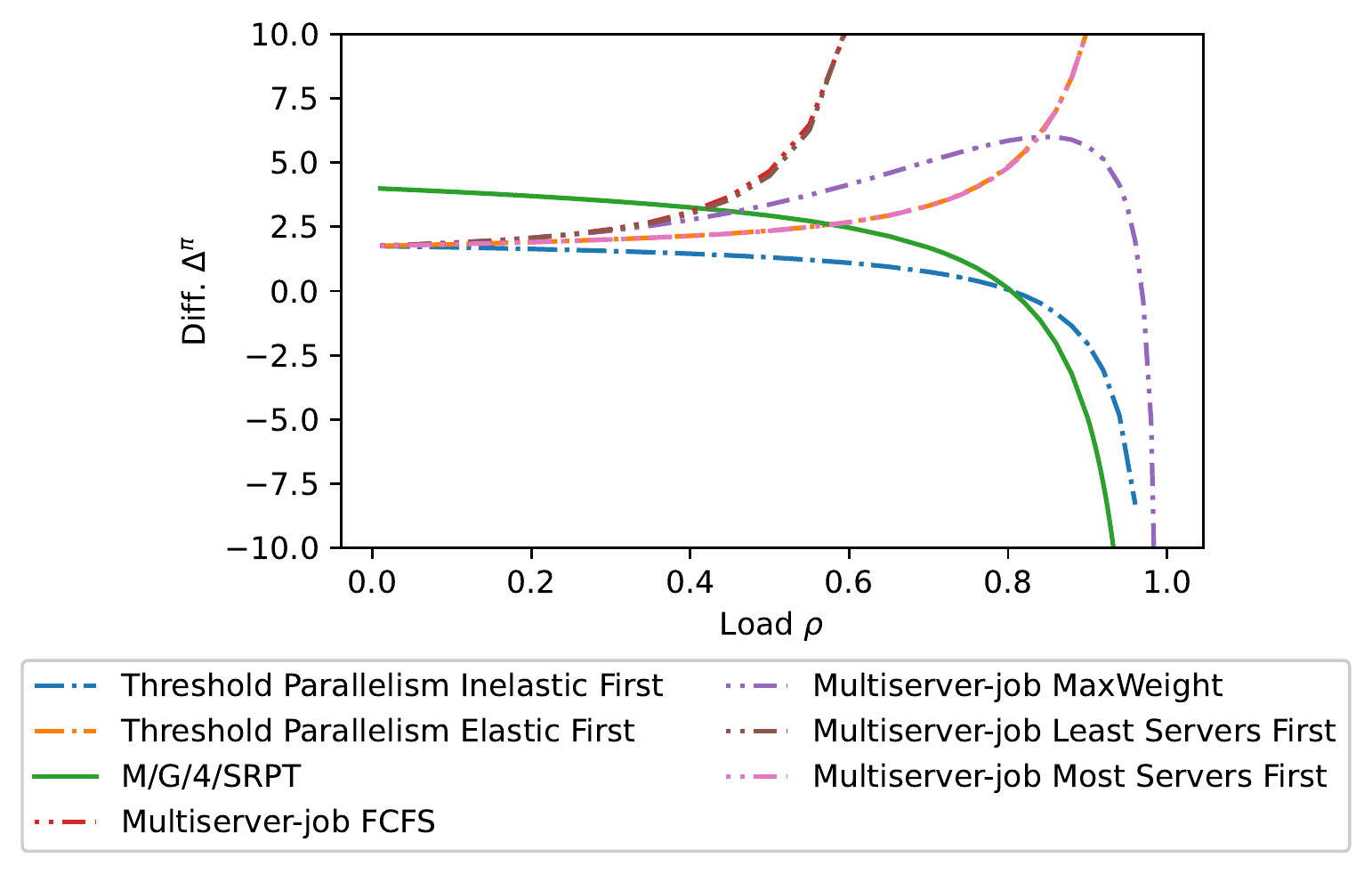}
    \caption{$\Delta^\pi$ for non-WCFS models.
    Same job sizes and specific settings as in \cref{fig:diff-wcfs}.
    Same number of arrivals and range of $\rho$
    except MaxWeight: $10^{10}$ arrivals, $\rho \in [0, 0.99]$.
    }
    \label{fig:diff-non-wcfs}
\end{figure}
In contrast, in \cref{fig:diff-non-wcfs},
we simulate several non-WCFS models,
which we depicted earlier in
\cref{fig:intro-response-other}.
These models are:
\begin{itemize}
    \item \textbf{Threshold Parallelism Inelastic First:}
    This is the Threshold Parallelism model from \cref{sec:threshold-parallelism},
    but rather than serving jobs in FCFS order,
    we prioritize jobs $j$ with smaller parallelism threshold $p_j$
    \cite{berg_towards_2017}.
    \item \textbf{Threshold Parallelism Elastic First:}
    This is the Threshold Parallelism model from \cref{sec:threshold-parallelism},
    but we prioritize jobs $j$ with larger parallelism threshold $p_j$.
    \item \textbf{M/G/k/SRPT:}
    This is an M/G/k,
    where each of the $k$ servers runs at speed $1/k$,
    and we prioritize
    jobs of least remaining size.
    \item \textbf{Multiserver-job FCFS:}
    This is the Multiserver-job model
    from \cref{sec:multiserver-jobs},
    but we serve jobs in FCFS order.
    If the next job to be served doesn't ``fit''
    in the remaining servers,
    those servers remain idle until other jobs complete,
    idling sufficient servers to allow
    the job to fit.
    \item \textbf{Multiserver-job Least Servers First:}
    This is the Multiserver-job model
    from \cref{sec:multiserver-jobs},
    but we prioritize jobs $j$ with smaller server requirements $v_j$.
    Again, if the next job doesn't fit,
    the remaining servers remain idle until the job can fit.
    \item \textbf{Multiserver-job Most Servers First:}
    This is the Multiserver-job model
    from \cref{sec:multiserver-jobs},
    but we prioritize jobs $j$ with larger server requirements $v_j$.
    \item \textbf{Multiserver-job MaxWeight:}
    This is the Multiserver-job model
    from \cref{sec:multiserver-jobs},
    but we serve jobs according to the ``MaxWeight''
    policy
    which we describe in \cref{sec:max-weight}.
\end{itemize}
In all cases, prioritization is preemptive.

Our empirical results in \cref{fig:diff-non-wcfs}
indicate that for these non-WCFS policies,
$\Delta^\pi$ diverges as $\rho \to 1$.
Specifically, for Threshold Parallelism Elastic First,
Multiserver-job FCFS,
Multiserver-job Least Servers First, and
Multiserver-job Most Servers First,
$\Delta^\pi$ appears to diverge in the positive direction.
For Threshold Parallelism Inelastic First,
M/G/k/SRPT,
and Multiserver-job ServerFilling,
$\Delta^\pi$ appears to diverge in the negative direction.
Note the expanded scale of \cref{fig:diff-non-wcfs}
as compared to \cref{fig:diff-wcfs}.
For
Multiserver-job MaxWeight,
we performed additional simulation,
which indicated that $\Delta^\pi$ diverged in the negative direction as $\rho \to 1$.

Next,
we explore the behavior of $\Delta^\pi$
for WCFS models,
as we vary the front size $n$
and the job size distribution $S$.

First, in \cref{fig:many-k},
we investigate the effects of varying front size $n$
on $\Delta^\pi$
for the Multiserver-job model with our ServerFilling policy;
under this model, the front size $n$ is equal to the number of servers $k$.
In this setting, the difference $\Delta^\pi$
empirically grows approximately linearly with the number of servers $k$,
and is nearly constant as $\rho \to 1$.
This matches the behavior of our bounds proven in \cref{thm:explicit-bounds},
which expand linearly with $n$.
Our simulations indicate that other WCFS policies 
similarly experience linear relationships between $n$ and $\Delta^\pi$.

In \cref{fig:many-s}
we investigate the effects of varying job size distribution $S$
on $\Delta^\pi$
in the Heterogeneous M/G/k where the
job size distribution $S$ is parameterized by a real value $x$.
Each $S$ is a hyperexponential distribution with $E[S] = 1$.
At large ages $a$,
the remaining size distributions $[S - a \mid S > a]$ of these job size distributions
converge to $Exp(1/x)$, the larger exponential branch.
From this, it is straightforward to show that
$\rem_{\sup}(S) = x$.

In \cref{fig:many-s},
we see that as $x$ increases, 
$\Delta^\pi$ at loads near 1 falls linearly,
with more negative slope for larger $x$.
However, for each specific $x$,
it does not appear that $\Delta^\pi$ is diverging to positive or negative infinity.
For instance, consider the red curve, $x=8$:
as $\rho \to 1$,
$\Delta^\pi$ converges to a value near $-3$, rather than diverging.

Broadly, \cref{fig:many-s}
matches the behavior of our bounds proven in \cref{thm:explicit-bounds},
which expand linearly with $\rem_{\sup}(S)$, which here is $x$.
We have empirically found that other WCFS policies similarly experience linear relations between
$\rem_{\sup}(S)$ and $\Delta^\pi$,
for hyperexponential job size distributions $S$,
and we believe that similar behavior will occur for other common job size distributions.

\section{Conclusion}

We introduce the \emph{work-conserving finite-skip} (WCFS) framework,
and use it to analyze many important queueing models which have eluded analysis thus far.
We prove that the scaled mean response time $E[T^\pi](1-\rho)$
of any WCFS model $\pi$ converges in heavy traffic to
the same limit as M/G/1/FCFS.
Moreover, we prove that the additive gap $\Delta^\pi = E[T^\pi] - E[T_Q^{M/G/1}]$
remains bounded by explicit constants at all loads $\rho$,
proving rapid convergence to the heavy traffic limit.

A possible direction for future work
would be to to tighten the explicit constants on $\Delta^\pi$.
Doing so will likely require use of more detailed properties of the WCFS models
being analyzed, but seems quite doable.

This paper considers models which are finite skip
and work conserving relative to the FCFS service ordering.
Another interesting direction would be to investigate policies which are
``finite-skip'' relative to other base service orderings.
Hopefully, one could prove bounds on mean response time of models in this new class
relative to an M/G/1 operating under the base service ordering.

Finally, one could try to characterize other metrics of response time for WCFS policies,
such as tail metrics.
One possible approach to doing so would be to generalize
the rate-conservation technique used in \cref{lem:work-bounds}.

\bibliography{refs}
\iffull
\begin{appendices}

\section{DivisorFilling}
\label{app:divisor-filling}

The DivisorFilling policy is a Multiserver-job service policy
which assumes that all server requirements $v_j$
divide the total number of servers $k$.
The DivisorFilling policy is a WCFS policy with front size $n = k$,
as we will show.
Finite-skip will be straightforward, the main difficulty is showing work-conservation.

We first define the DivisorFilling policy.
DivisorFilling is a preemptive policy,
in that when a job completes, the set of jobs in service may change,
removing partially-complete jobs from service.
The DivisorFilling policy is defined recursively.
The policy's behavior with respect to larger $k$
is defined based on its behavior for smaller $k$.
In particular, we will prove work conservation inductively.

Let $M$ be the set of jobs at the front.

To define DivisorFilling, we split into three cases:
\begin{itemize}
    \item $M$ contains at least $k/6$ jobs with server requirement $v_j = 1$.
    \item $k = 2^a3^b$ for some integers $a, b$,
    and $M$ contains $< k/6$ jobs with $v_j = 1$.
    \item $k$ has a prime factor $p \ge 5$
    and $M$ contains $< k/6$ jobs with $v_j = 1$.
\end{itemize}

\subsection{At least $k/6$ jobs requiring 1 server}
First, assume that $M$ contain at least $k/6$ jobs requiring 1 server.

Just as in the ServerFilling policy,
label the jobs $f_1, f_2, \ldots$ in decreasing order of server requirement.
Let $i^*$ be defined as
\begin{align*}
    i^* = \arg \max_i \sum_{\ell = 1}^i v_{f_\ell} \le k.
\end{align*}

In this case, the DivisorFilling policy
serves jobs $f_1, \ldots f_{i^*}$,
as well as any jobs requiring 1 server that fit in the remaining servers.
Specifically,
DivisorFilling serves
\begin{align*}
    k - \sum_{\ell = 1}^{i^*} v_{f_\ell}.
\end{align*}
additional jobs that require 1 servers,
or all jobs requiring 1 server if fewer are available.

\subsubsection{Work conservation}
We want to show that if $M$ contains $k$ jobs,
DivisorFilling serves jobs requiring $k$ servers
in this case.

Let us write $\textsc{sum}_{i^*} := \sum_{\ell = 1}^{i^*} v_{f_\ell}$.
Because we have at least $k/6$ jobs requiring 1 server,
it suffices to show that $\textsc{sum}_{i^*} \ge 5k/6$.
The remaining servers are filled by the jobs requiring 1 server.

First, note that $\textsc{sum}_k \ge k$,
because there are $k$ jobs, each requiring at least 1 server.
Next, note that $k - \textsc{sum}_{i^*} < f_{i^* + 1}$,
because the $i^*+1$ job does not fit in service.
Because the labels $f_1, f_2, \ldots$ are in decreasing order of server requirement,
$k - \textsc{sum}_{i^*} < f_{i^*}$.

Therefore, to prove that $k - \textsc{sum}_{i^*} \le k/6$,
we need only consider sequences of the $i^*$ largest server requirements in $M$
in which all such requirements
are greater than $k/6$.
We need only consider requirements equal to $k, k/2, k/3, k/4, k/5$.

We enumerate all such sequences.
Note that if $k$ is not divisible by all of $\{2, 3, 4, 5\}$,
some entries will not apply.
This only tightens the resulting bound on $k - \textsc{sum}_{i^*}$ for such $k$.

We list $i^*$ requirements if $\textsc{sum}_{i^*} = k$,
and $i^*+1$ otherwise.
We write $g_{i^*}$ as a shorthand for $k - \textsc{sum}_{i^*}$.

\begin{center}
\begin{tabular}{| l | l | l | l |}
\hline
Sequence & $g_{i^*}$ & Sequence & $g_{i^*}$ \\
\hline
$k $&$ 0 $&$ k/2, k/2 $&$ 0 $\\
$k/2, k/3, k/3 $&$ k/6 $&$ k/2, k/4, k/4 $&$ 0 $\\
$k/2, k/4, k/5, k/5 $&$ k/20 $&$ k/2, k/5, k/5, k/5 $&$ k/10 $\\
$k/3, k/3, k/3 $&$ 0 $&$ k/3, k/3, k/4, k/4 $&$ k/12 $\\
$k/3, k/3, k/5, k/5 $&$ 2k/15 $&$ k/3, k/4, k/4, k/4 $&$ k/6 $\\
$k/3, k/4, k/5, k/5, k/5 $&$ k/60 $&$ k/3, k/5, k/5, k/5, k/5 $&$ k/15 $\\
$k/4, k/4, k/4, k/4 $&$ 0 $&$ k/4, k/4, k/4, k/5, k/5 $&$ k/20 $\\
$k/4, k/4, k/5, k/5, k/5 $&$ k/10 $&$ k/4, k/5, k/5, k/5, k/5 $&$ 3k/20 $\\
$k/5, k/5, k/5, k/5, k/5 $&$ 0 $&&\\
\hline
\end{tabular}
\end{center}

In all cases, $k - \textsc{sum}_{i^*} \le k/6$.
As a result, DivisorFilling is work conserving in this case.

\subsection{$k = 2^a3^b$}
\label{sec:2-and-3}
Suppose that $k$ is of the form $2^a3^b$, for some integers $a$ and $b$,
and that the number of jobs in $M$
that require 1 server is less than $k/6$.

Let $M_2$ be the set of jobs requiring an even number of servers in $M$,
and let $M_r$ be the remaining jobs:
\begin{align*}
    M_2 &:= \{j \mid j \in M, v_j \text{ is even}\} \\
    M_r &:= \{j \mid j \in M, v_j \text{ is odd}, v_j > 1\}
\end{align*}

Note that because 2 and 3 are the only prime factors of $k$,
all jobs in $M_r$ have server requirements divisible by 3.

How we now schedule is based on which is larger: $2\vert M_2\vert $,
or $3\vert M_r\vert $. In this case of a tie,
either would be fine, so we arbitrarily select $M_2$.

If $2\vert M_2\vert $ is larger,
we will only serve jobs from among $M_2$.
To do so, imagine that we combine pairs of servers,
reducing $k$ by a factor of 2, and reducing
the server requirement of every job in $M_2$ by a factor of 2.
We now compute which jobs from $M_2$ DivisorFilling would serve,
in this simplified subproblem.
DivisorFilling serves the corresponding jobs.

If $3\vert M_r\vert $ is larger,
we do the same, except that we combine triples of jobs.

\subsubsection{Work conservation}
If at least $k$ jobs are present, we will show that this process fills all of the servers.

Because there are $< n/6$ jobs requiring 1 server, $\vert M_2\vert  + \vert M_3\vert  \ge 5k/6$.
As a result, either $2\vert M_2\vert  \ge k$ or $3\vert M_r\vert  \ge k$.
Consider the case where $2\vert M_2\vert  \ge k$.
The constructed subproblem has $k/2$ servers and $\vert M_2\vert  \ge k/2$
jobs, so by induction DivisorFilling fills all of the servers in the subproblem.
That property is carried over in the main problem.
The case where $3\vert M_r\vert  \ge k$ is equivalent.

\subsection{$k$ has a prime factor $k \ge 5$}
Finally, suppose that $k$ has a prime factor $p \ge 5$,
and that $M$ contains $< k/6$ jobs requiring 1 server.
Specifically, let $p$ be $k$'s largest prime factor.

Let us form the set $M_p$
consisting of the jobs in $M$ whose server requirements are multiples of $p$,
and $M_r$ consisting of jobs which require more than 1 server, but not a multiple of $p$.
As in \cref{sec:2-and-3},
if $\vert M_p\vert  \ge k/p$, we can recurse by combining groups of $p$
servers to fill all of $M$.

Otherwise, we turn to $M_r$.
Note that all jobs in $M_r$ have server requirements which are divisors of $k/p$,
because their requirements are divisors of $k$ which are not multiples of $p$.

If $\vert M_r\vert  \ge k/p$,
let us apply the DivisorFilling policy
on an arbitrary subset of $M_r$ of size $k/p$.
By induction, 
DivisorFilling finds a subset of these jobs requiring exactly $k/p$ servers.
Let us extract this subset from $M_r$, creating $M_r^1$.
We repeat this process until we have extracted $p$ subsets,
or $\vert M_r^i\vert  < k/p$ for some $i$.
DivisorFilling serves the extracted subsets.

\subsubsection{Work conservation}
We must show that the extraction procedure always successfully extracts $p$ subsets,
if $\vert M\vert  = k$.

In the extraction case, note that $\vert M_p\vert  < k/p \le k/5$,
and that there are $\le k/6$ jobs requiring 1 server.
$M_r$ consists of the remaining jobs.
As a result,
\begin{align*}
    \vert M_r \vert \ge k - k/6 - k/5 = 19k/30.
\end{align*}

Note also that every job in $M_r$ requires at least 2 servers,
so at most $k/2p$ jobs are extracted at each step.
To prove that $p$ subsets can be extracted,
we must show that at least $k/p$ jobs remain after $p-1$ subsets have been extracted.
\begin{align*}
    \vert M_r^{p-1} \vert
    \ge \frac{19k}{30} - \frac{(p-1)k}{2p} = \frac{19k}{30} - \frac{k}{2} + \frac{k}{2p}
    = \frac{2k}{15} + \frac{k}{2p}
\end{align*}
To prove that $\vert M_r^{p-1} \vert \ge k/p$,
we just need to show that $2k/15 \ge k/2p$.
But $p \ge 5$, so $2k/15 > k/10 \ge k/2p$.

Thus, we can always extract $p$ disjoint subsets of jobs,
each requiring a total of $k/p$ servers,
from $M_r$. Combining these subsets fills all $k$ servers, as desired.
\end{appendices}

\fi
\end{document}